\documentclass[twocolumn, 10pt]{IEEEtran}

\usepackage{hyperref} 
\usepackage{graphicx}
\usepackage{amssymb}
\usepackage{amsmath}
\usepackage{mathtools}
\usepackage{cite}
\usepackage{stfloats}
\usepackage{subfigure}
\usepackage{psfrag}
\usepackage[mathscr]{euscript}
\usepackage{acronym}  
\usepackage{booktabs} 
\usepackage[table]{xcolor}
\usepackage{color}
\usepackage[utf8]{inputenc}

\usepackage{comment}
\usepackage{caption}
\DeclareCaptionLabelFormat{algnonumber}{Algorithm}
\usepackage{algorithm}
\usepackage{algpseudocode}
\algtext*{EndFor}
\algtext*{EndWhile}
\algtext*{EndIf}

\acrodef{BS}{base station}
\acrodef{MEC}{mobile edge computing}
\acrodef{VUE}{vehicular user equipment}
\acrodef{LoS}{Line-of-Sight}
\acrodef{NLoS}{Non-Line-of-Sight}
\acrodef{UDN}{Ultra-Dense Network}
\acrodef{WSN}{wireless sensor network}

\acrodef{PPP}{Poisson point process}
\acrodefplural{PPP}[PPPs]{Poisson point processes}

\acrodef{PDF}{probability density function}
\acrodef{PMF}{probability mass function}
\acrodef{CDF}{cumulative distribution function}
\acrodef{CCDF}{complementary cumulative distribution function}
\acrodef{SIR}{signal-to-interference ratio}
\acrodef{SINR}{signal-to-interference-plus-noise ratio}
\acrodef{SNR}{signal-to-noise ratio}
\acrodef{PGFL}{probability generating functional}
\acrodef{ASE}{area spectral efficiency}

\acrodef{V2X}{vehicle-to-everything}
\acrodef{RGM}{random Gauss-Markov}
\acrodef{STP}{successful transmission probability}
\acrodef{RSU}{road side unit}
\acrodef{VUE}{vehicular user equipment}

\acrodef{P-K}{Pollaczek-Khinchin}
\acrodef{KKT}{Karush-Kuhn-Tucker}

\acrodef{MISO}{multi-input single-output}
\acrodef{MRT}{maximal ratio transmission}

\acrodef{IoT}{Internet of things}
\acrodef{RF}{radio frequency}
\acrodef{AI}{artificial intelligence}
\acrodef{AoI}{age of information}
\acrodef{WSN}{wireless sensor networks}
\acrodef{EH}{energy harvesting}
\acrodef{EH-WSN}{energy harvesting wireless sensor network}
\acrodef{EC}{edge computing}
\acrodef{LC}{local computing}
\acrodef{MARL}{multi-agent reinforcement learning}
\acrodef{MADRL}{multi-agent deep reinforcement learning}
\acrodef{RL}{reinforcement learning}
\acrodef{MDP}{Markov Decision Process}
\acrodef{POMDP}{partially observable Markov decision process}
\acrodef{DPG}{deterministic policy gradient}
\acrodef{DDPG}{Deep deterministic policy gradient}
\acrodef{MADDPG}{multi-agent deep deterministic policy gradient}
\acrodef{DTDE}{decentralized training with decentralized execution}
\acrodef{CTDE}{centralized training with decentralized execution}
\acrodef{IS}{importance sampling}
\acrodef{SD}{sensing decision}
\acrodef{SCD}{sensing and computing decision}
\acrodef{JoSCDF}{Joint Sensing and Computing decision for Data Freshenss}
\acrodef{AWGN}{Additive White Gaussian Noise}
\acrodef{SIC}{successive interference cancellation}

\usepackage{color}
\usepackage{dsfont}
\usepackage{bbm}

\newcommand{\argmax}{\mathop{\mathrm{argmax}}}

\DeclareMathAlphabet{\mathsf}{OML}{cmbr}{m}{it}

\newtheorem{lemma}{Lemma}

\newtheorem{remark}{Remark}

\newcommand{\loc}{\ell}
\newcommand{\edge}{\text{e}}
\newcommand{\SensorSet}{\mathcal{N}}
\newcommand{\SensorNum}{N_\text{s}}

\newcommand{\Ptx}[1]{P_{\text{tx}}}
\newcommand{\InDataSize}{L_{\edge}}
\newcommand{\OutDataSize}{L_{\loc}}

\newcommand{\ProcSlotE}{\tau_{\edge}}
\newcommand{\ProcSlotL}{\tau_{\loc}}
\newcommand{\SNR}[1]{\zeta_{\textbf{x}_{#1},\text{o}}}
\newcommand{\Bandwidth}{W}
\newcommand{\ChannelFading}[1]{h_{\textbf{x}_{#1},\text{o}}}
\newcommand{\DistToSink}[1]{d_{\textbf{x}_{#1},\text{o}}}
\newcommand{\DistToPoint}[1]{d_{\textbf{x}_{#1},\textbf{y}}}
\newcommand{\PathExp}{\alpha}

\newcommand{\OutageProb}[1]{p_{\textbf{o},{#1}}}

\newcommand{\Txcount}{\delta}
\newcommand{\PreBatteryTh}{B_{\text{th}}}
\newcommand{\Bmax}{B_{\text{max}}}
\newcommand{\EnergyLevel}[2]{B_{#1}(#2)}
\newcommand{\ConsumeEnergy}[2]{b_{#1}[#2]}
\newcommand{\HarvEnergy}[2]{\varepsilon_{#1}[#2]}  

\newcommand{\SensingRad}{r_{\text{c}}}                          
\newcommand{\SlotDuration}{\tau_{\text{t}}}                           
\newcommand{\RoundLen}{\tau_{\text{r}}}
\newcommand{\TotRound}{N_{\text{r}}}
\newcommand{\RoundIdx}{j}
\newcommand{\RoundIdxTime}[1]{\tilde{t}_{#1}}

\newcommand{\SensorAoI}[2]{\Delta_{#1}^{\text{tx}}(#2)}
\newcommand{\SinkAoI}[2]{\Delta_{#1}^{\text{rx}}(#2)}
\newcommand{\SensorAoISgl}[1]{\Delta^{\text{tx}}(#1)}
\newcommand{\SinkAoISgl}[1]{\Delta^{\text{rx}}(#1)}
\newcommand{\WhichAoI}[2]{\Delta_{#1}^{\WhichGenUpdateTimeIdx}(#2)}
\newcommand{\GenTime}[2]{{#2}^\text{tx}_{\text{g}, #1}}
\newcommand{\UpdateTime}[2]{{#2}^\text{rx}_{\text{g}, #1}}
\newcommand{\WhichGenUpdateTimeIdx}{\iota}
\newcommand{\WhichGenUpdateTime}[2]{{#2}^{\WhichGenUpdateTimeIdx}_{\text{g}, #1}}
\newcommand{\Coordinate}{\Phi}
\newcommand{\CoordinateSingle}{\Phi_{\text{sg}}}

\newcommand{\Covariance}[1]{\rho_{\textbf{x}_{#1},\textbf{y}}}
\newcommand{\EstimationErr}[1]{\epsilon_{\textbf{x}_{#1},\textbf{y}}}
\newcommand{\ErrWeight}[1]{\beta_{#1}}
\newcommand{\ErrTh}{\epsilon_{0}}
\newcommand{\PointSensorCov}[1]{c_{\textbf{x}_{#1},\textbf{y}}}
\newcommand{\PointCov}{C_{\textbf{y}}}
\newcommand{\NetCov}{\phi_{\text{cov}}}

\newcommand{\TgtCov}{\eta}
\newcommand{\ECovProb}{\mathcal{P}_\text{c}(\TgtCov)}
\newcommand{\ECovProbPsPe}{\mathcal{P}_\text{c}^{\TgtCov}(\Ps,\Pe)}
\newcommand{\E}[1]{E_{#1}}
\newcommand{\SE}{\mathrm{S}}
\newcommand{\T}{\mathrm{T}}

\newcommand{\Harv}{\mathrm{H}}
\newcommand{\Ps}{p_{\text{s}}}
\newcommand{\Pe}{p_{\edge}}

\newcommand{\aratio}{$\TgtCov$-coverage probability}
\newcommand{\Aratio}{$\TgtCov$-Coverage Probability}
\newcommand{\AVP}{AoI violation probability}

\newcommand{\Uk}{\mathbb{E}[U_k]}
\newcommand{\Xk}{\mathbb{E}[X_k]}
\newcommand{\Zk}{\mathbb{E}[Z_k]}
\newcommand{\gk}{\mathbb{E}[g_k]}
\newcommand{\Peprime}[1]{p_{\text{u}}(#1)}
	
\newcommand{\tk}{\tau_{k}}
\newcommand{\tkmo}{\tau_{k-1}}
\newcommand{\y}{y}
\newcommand{\Y}{Y}

\newcommand{\TxProbPout}[2]{1 - (\OutageProb{#1})^{{#2}}}
\newcommand{\TxProbPf}[1]{\varrho_{\Txcount,#1}}
\newcommand{\TxProbxTime}[2]{\varrho_{#2,#1}}
\newcommand{\Px}{p_{\Txcount}}

\newcommand{\vth}{v_{\text{th}}}
\newcommand{\vfloor}{\lfloor \vth \rfloor}

\newcommand{\AvgTxNum}[2]{\bar{N}_{#1}(\OutageProb{#2})}

\newcommand{\AVPCaseOne}{\vth < Z_{k-1}}
\newcommand{\AVPCaseTwo}{Z_{k-1} \leq \vth < \RoundLen + X_k}
\newcommand{\AVPCaseThree}{\RoundLen + X_k \leq \vth}

\newcommand{\FuncOne}[2]{f_{#2,#1}}
\newcommand{\FuncTwo}[3]{h_{#3,#1}({#2})}
\newcommand{\FuncThree}[2]{\kappa_{#1}(#2)}

\newcommand{\POne}{\hat{P}_1 (\vth)}
\newcommand{\PTwo}{\hat{P}_2 (\vth)}
\newcommand{\PThree}{\hat{P}_3 (\vth)}

\newcommand{\UpdateKMO}{m_{k-1}}
\newcommand{\UpdateK}{m_{k}}

\newcommand{\Es}{E_{\text{S}}}
\newcommand{\Et}{E_{\text{T}}}
\newcommand{\Ep}{E_{\text{C}}}

\newcommand{\RewardPenalty}{\nu}

\newcommand{\ObsRange}{d_\text{obs}}
\newcommand{\Softupdate}{\chi}
\newcommand{\HiddenNodeNum}[1]{H_{#1}}
\newcommand{\NLayerActor}{L_A}
\newcommand{\NLayerCritic}{L_C}
\newcommand{\SumI}{I_{n}}
\newcommand{\LaplaceTrans}[2]{\mathcal{L}_{#1}\left({#2}\right)}
\newcommand{\BSdensity}{\lambda}
\newcommand{\Idensity}{\lambda_{\text{I}}}

\newcommand{\bd}{\begin{description}}
\newcommand{\ed}{\end{description}}
\newcommand{\be}{\begin{enumerate}}
\newcommand{\ee}{\end{enumerate}}
\newcommand{\bi}{\begin{itemize}}
\newcommand{\ei}{\end{itemize}}
\newcommand{\bl}{\begin{list}}
\newcommand{\el}{\end{list}}
\newcommand{\bt}{\begin{tabbing}}
\newcommand{\et}{\end{tabbing}}

\newcolumntype{C}[1]{>{\centering\arraybackslash}p{#1}}

\newcommand{\blue}[1]{{#1}}

\begin{document}
	
	\newcommand{\paperTitle}{Learning-based Sensing and Computing Decision
		\\for Data Freshness\\in Edge Computing-enabled Networks}
	
	\title{\paperTitle}
	\author{Sinwoong Yun, Dongsun Kim, Chanwon Park, and Jemin Lee, \textit{Member, IEEE}
		
		\thanks{
			Corresponding author is J. Lee.
		}
		\thanks{
			S. Yun and D. Kim are with the Department of Electrical Engineering and Computer Science, Daegu Gyeongbuk Institute of Science and Technology (DGIST),
			Daegu 42988, Republic of Korea (e-mail: \{lion4656, yidaever\}@dgist.ac.kr).
		}
		\thanks{
			C. Park is with the Agency for Defense Development (ADD), Daejeon 34186, Republic of Korea (e-mail: chwpark91@gmail.com).
		}
		\thanks{
			J. Lee is with the School of Electrical and Electronic Engineering, Yonsei University, Seoul 03722, South Korea (e-mail: jemin.lee@yonsei.ac.kr).
		}
		\thanks{This article was presented in part at the IEEE Global Communications Conference, Malaysia, December 2023 \cite{YunKimPar:23}.
		}
		\thanks{This work has been submitted to the IEEE for possible publication. Copyright may be transferred without notice, after which this version may no longer be accessible.
	}
	}
	\maketitle 
	
	\setcounter{page}{1}
	\acresetall
	\vspace{-22mm}
	\begin{abstract}
		As the demand on \ac{AI}-based applications increases, the freshness of sensed data becomes crucial in the wireless sensor networks.
		Since those applications require a large amount of computation for processing the sensed data, it is essential to offload the computation load to the \ac{EC} server. 
		In this paper, we propose the \ac{SCD} algorithms for data freshness in the \ac{EC}-enabled wireless sensor networks.
		We define the \aratio\ to show the probability of maintaining fresh data for more than $\TgtCov$ ratio of the network, where the spatial-temporal correlation of information is considered.
		We then propose the probability-based \ac{SCD} for the single pre-charged sensor case with providing the optimal point after deriving the \aratio.
		We also propose the \ac{RL}-based \ac{SCD} by training the \ac{SCD} policy of sensors for both the single pre-charged and multiple \ac{EH} sensor cases, to make a real-time decision based on its observation.
		Our simulation results verify the performance of the proposed algorithms under various \blue{environment settings}, and show that the \ac{RL}-based \ac{SCD} algorithm achieves higher performance compared to baseline algorithms for both the single pre-charged sensor and multiple \ac{EH} sensor cases.
	\end{abstract} 
	\acresetall
	
	\section{Introduction}\label{sec:intro} 
	The new generation of wireless networks will realize the intelligence of everything,
	which integrates the digital world and the physical world by sensing everything \cite{VerFri:13}.
	This will lead to the change on the role of \ac{WSN}
	from providing sensed data efficiently  
	to providing \emph{fresh} data reliably. 
	The data freshness is especially crucial 
	for \ac{AI}-based decision making applications and services 
	such as the autonomous driving and the security surveillance \cite{YicMukGho:08}.
	In those services, the outdated information can cause the fatal problems 
	such as the car accident or the invasion. 
	For maintaining the data freshness in the sensor networks, 
	the sensor should be continuously activated and transmit the sensed data to the sink node that manages the data. 
	However, because of the limited battery capacity of sensors, 
	frequent sensing and transmission may shorten the sensor lifetime and fail to maintain the data freshness in the networks.
	Hence, both the energy consumption and the data freshness should be considered in the design of sensor networks.
	%

	Furthermore, as the demand on \ac{AI}-based applications and services is increasing, 
	the processing of the sensed data requires a larger amount of computation. 
	The large computation might not be able to be done at the sensors 
	due to the large energy consumption as well as the long processing delay from the low computing capability of sensors. 
	Hence, the \ac{EC}, which is to offload the computation load of sensors to the \ac{EC} server, becomes essential for energy-efficient wireless sensor networks \cite{ShiCaoZha:16, MaoYouJha:17}.
	

	Recently, the data freshness has been considered in many works for the energy-efficient sensor networks\cite{LiuZhoDur:16, Kri:19, BacUys:17, WuYanWu:17, CerGunGyo:19, MolLeiCod:20, KadSinMod:18, HriCoKamDa:18,HriMarChi:20}.
	Here, to measure the data freshness, 
	the \ac{AoI} has been used, which is the elapsed time since the generation of the data\cite{KauYatGru:12}.
	For the single sensor case, the average \ac{AoI} \cite{Kri:19} 
	and the \ac{PMF} of \ac{AoI} \cite{LiuZhoDur:16} are studied for given sensing decision period. 
	The sensing decision algorithms are presented to minimize the average \ac{AoI} of the single sensor 
	using the threshold-based \cite{BacUys:17, WuYanWu:17} and the \ac{RL}-based approach \cite{CerGunGyo:19}.
	%
	%
	By considering multiple sensors,
	the average \ac{AoI} is analyzed in \cite{MolLeiCod:20},
	when the sensing instances are determined randomly and independently  
	with the exponentially-distributed intervals.  
	The sensing decision algorithms are also presented to minimize the average \ac{AoI} of sensors
	using the Lyapunov optimization and the \ac{RL}-based approach\cite{KadSinMod:18}.
	However, in those works, the \ac{EC} is not considered in the design of the sensor networks.

	
	The \ac{EC}-enabled sensor networks have been introduced in recent works \cite{XuYeYan:19, JosDan:20}. 
	In those works, the computing decision has been studied 
	to determine whether the sensor computes the task locally or offloads it to the \ac{EC} server.
	Specifically, for multiple sensor case, the computing decision algorithms are presented 
	to minimize the latency and the computing energy consumption 
	using the convex optimization-based \cite{XuYeYan:19} and the game theory-based approach \cite{JosDan:20}.
	Recently, in \cite{MaZhoSun:21}, the data freshness is also considered in the computing decision algorithm, which is designed to minimize the weighted sum of computing delay and energy consumption with guaranteeing the data freshness constraint.
	As such, most existing studies are focused on either sensing or computing decision.
	Nevertheless, since both the sensing decision and computing decision affect the performance of the wireless sensor networks, they should be jointly designed.
	

	
	The joint \ac{SCD} has been designed for data freshness in \cite{LiMaGon:21,XuYanWan:19}.
	For the single sensor case, the \ac{RL}-based \ac{SCD} algorithm is proposed to minimize the average \ac{AoI}.
	Moreover, for the multiple sensor case, the Lyapunov optimization-based \ac{SCD} algorithm is presented to minimize the peak \ac{AoI}.
	%
	Generally, the sensed data has a temporal correlation, which means it can be similar to the data, sensed in a short time ago \cite{TayMakPer:19}. 
	Furthermore, in the wireless sensor networks, where sensors are densely deployed, 
	the sensing coverage of nearby sensors can have overlapping areas, which means the sensed data from the sensors can have a spatial correlation as well \cite{WanDenLiu:13}.
	Those correlation affect the accuracy of the sensed data, which decreases over time and as the distance to the sensing point increases \cite{HriCoKamDa:18}.
	Therefore, in the energy-efficient design of the \ac{SCD} for the multiple sensor case, 
	the temporal and the spatial correlation of sensed data should be jointly considered, but in  \cite{LiMaGon:21,XuYanWan:19}, only the temporal correlation is considered. 
	
	%

%
%

Thus, in this paper, we propose the \ac{SCD} algorithms for data freshness in the \ac{EC}-enabled wireless sensor networks by considering the spatial-temporal correlation of information. 
Specifically, we consider that the information at different locations, estimated from the sensed data, 
has the error, which increases with the distance from the sensor and the elapsed time from sensing.
Therefore, we define the error-tolerable network coverage ratio as the portion of network area with smaller error than certain threshold.
We then propose the \aratio\ as a performance metric, which is the probability that the network coverage ratio is greater than the target value $\TgtCov$.
To maximize the \aratio, we propose the \ac{SCD} algorithms for two cases: the single pre-charged sensor case and the multiple \ac{EH} sensor case, respectively.
We first propose the probability-based \ac{SCD} algorithm, where \blue{each sensor makes a decision based on} the sensing and \ac{EC} probabilities.
Next, we propose the \ac{RL}-based \ac{SCD} algorithm, where each sensor takes an action based on the real-time information as an observation.
The main contributions of this paper can be summarized as follows.
\begin{itemize}
	\item We propose the \ac{SCD} algorithms for data freshness in the \ac{EC}-enabled wireless sensor networks.
	To the best of our knowledge, this is the first work that jointly decides the sensing and computation considering both the spatial and the temporal correlation.
	
	\item In the probability-based \ac{SCD} algorithm, we derive the \aratio\ and provide the optimal sensing probability that maximizes the \aratio\ in a closed form for the single pre-charged sensor case. 
	We also provide the method to obtain the optimal sensing and \ac{EC} probabilities for the multiple \ac{EH} sensor case. 
	
	\item In the \ac{RL}-based \ac{SCD} algorithm, we train the policy for each sensor to make the \ac{SCD} according to the dynamic change of environments including the battery level of the sensors and the channel fading gain for the multiple EH sensor case as well as the single pre-charged sensor case.
	
	\item In simulation results, we verify the \aratio\ of the proposed algorithms for various \blue{environment settings} such as the target network coverage ratio, the distance between the sensor and the sink node, the number of sensors, and the computing energy.

\end{itemize}
\begin{figure}[t!]
	\begin{center}   
		{ 
			\includegraphics[width=1\columnwidth]{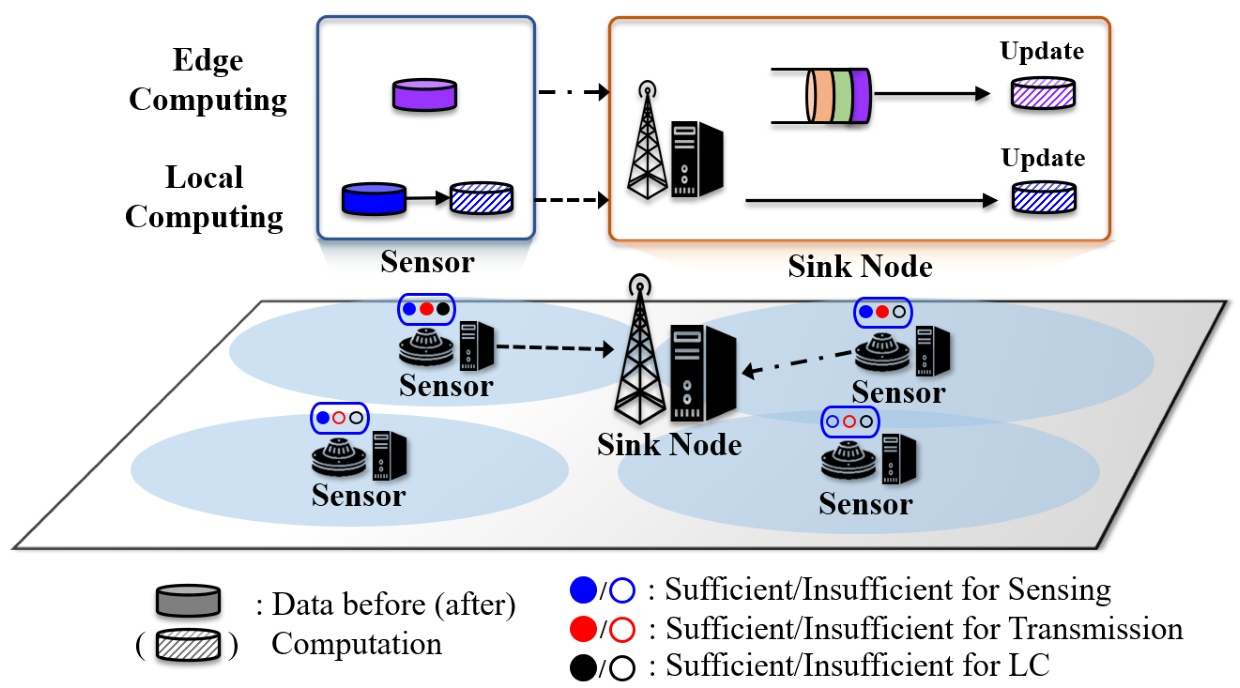}
		}
	\end{center}
	\caption{
		Network description of \ac{EC}-enabled wireless sensor networks.
	}
	\label{fig:System_model}
\end{figure}

\blue{
	The remainder of this paper is outlined as follows. 
	In Section \ref{sec:Network_model}, we firstly describe the \ac{EC}-enabled wireless sensor networks model, and define the \aratio\ as a performance metric by introducing the \ac{AoI}.
	To maximize the \aratio\ for both the single pre-charged and the multiple \ac{EH} sensor cases, we propose two algorithms, the probability-based \ac{SCD} in Section \ref{sec:Probability_SCD} and the \ac{RL}-based \ac{SCD} in Section \ref{sec:MARL_SCD}.
	In Section \ref{sec:Result}, the simulation results are presented and compared with baselines. 
	Finally, the conclusion of this paper is given in Section \ref{sec:Conclusion}.  
}


\section{Edge Computing-enabled Wireless Sensor Networks Model}\label{sec:Network_model}
In this section, we describe the \ac{EC}-enabled wireless sensor networks model with the transmission and energy models of the networks.
After that, we define the \aratio\ by introducing the \ac{AoI}.

\subsection{Network Description}\label{ssec:Network_model}
We consider the \ac{EC}-enabled wireless sensor networks, operating in a time-slotted fashion, which are composed of sink nodes distributed according to a homogeneous \ac{PPP} with intensity $\BSdensity$.\footnote{\blue{Note that we have employed the \ac{PPP} model to consider the realistic node deployment \cite{HaeGan:09} as well as the realistic distribution of uplink interference.}}
Each sink node equipped with the \ac{EC} server is associated with nearby $\SensorNum$ sensors and collects the data from them.
An example of the sink node and the associated sensors is given in Fig. \ref{fig:System_model}.
We use $\SensorSet = \{1,2,\cdots,\SensorNum\}$ as the index set of sensors associated with the sink node, and they have limited battery capacity.

Each sensor makes the \ac{SCD} at the beginning of each round, which consists of $\RoundLen$ time slots.
If the sensor performs sensing, it collects the data with size $\InDataSize$ from the surrounding area (e.g., temperature and humidity) for one time slot with consuming the energy $\E{\SE}$.
Then, the sensed data needs to be computed and transmitted to the sink node.
Both the sensors and the \ac{EC} server have computing capability, so the sensed data can be locally computed at the sensor, i.e., \ac{LC}, or offloaded to the \ac{EC} server, i.e., \ac{EC}.
If the sensor chooses the \ac{LC}, the data is processed at the sensor for $\ProcSlotL$ time slots with consuming the energy $\Ep$, and it is transmitted to the sink node and the status of the sensor is updated.
On the other hand, if the sensor chooses the \ac{EC}, the data is transmitted to the sink node and computed at the \ac{EC} server. 
When the newly sensed data of a sensor arrives at the \ac{EC} server, but the pre-generated old data of the corresponding sensor still exists at the queue, the old data is replaced with the fresh data.\footnote{By processing the lately sensed data instead of the existing data in the queue, the sink node can obtain the latest data generated at the sensor which reduces the \ac{AoI} at the sink node (will be described in Section \ref{ssec:Network_coverage}).}
After the \ac{EC} server processes the data of the sensor for $\ProcSlotE$ time slots, the status of the sensor is updated.\footnote{When the data of the $n$-th sensor arrives at the \ac{EC} server, it might need to wait for the computing as there can be other computing jobs of other sensors, arrived earlier.}
Generally, the computing capability of \ac{EC} server is higher than that of the sensor, so $\ProcSlotE < \ProcSlotL$.
Note that after the computation, the data size reduces to $\OutDataSize$ where $\OutDataSize < \InDataSize$.\footnote{In general, the output data size after the computation is smaller than the input data size \cite{CheJiaLi:15}.}

\subsection{Transmission Model}\label{ssec:Transmission_model}

The $n$-th sensor located at $\textbf{x}_{n}$ transmits the data with the transmission power $\Ptx{n}$ to the sink node, located at $\text{o}$, with consuming the energy $\E{\T}$.
When the sink node receives the sensed data from the $n$-th sensor, the \ac{SINR} $\SNR{n}$ is given by 
\begin{align}\label{eq:SINR}
	\SNR{n} = \frac{\Ptx{n} \ChannelFading{n} \DistToSink{n}^{-\PathExp}}{\sigma^2 + \SumI},
\end{align}
where $\ChannelFading{n}$ is the channel fading gain, $\DistToSink{n}$ is the distance between the $n$-th sensor and the sink node, $\PathExp$ is the path loss exponent, $\sigma^{2}$ is the \ac{AWGN} power.
In \eqref{eq:SINR}, $\SumI$ is the inter-cell interference.
\blue{Each sink node} has one sensor that uses the same frequency band with \blue{the $n$-th sensor \cite{ElsHos:14}.}
\blue{The distribution of interfering sensors does not constitute a homogeneous \ac{PPP} due to the dependency of their locations to the sink nodes, but 
it is shown that this dependency is weak \cite{NovDhiAnd:13}.
Hence, we assume the distribution of sensors interfering the $n$-th sensor follows a \ac{PPP} with intensity $\BSdensity$.}
\blue{When the probability of using the same frequency resource in other sink nodes for sensors is $p_{\text{t}}$, the density of interfering nodes becomes $\Idensity = \BSdensity p_{\text{t}}$, and their distribution is a {PPP} from the thinning property \cite{BacBla:09}. 
Therefore, the inter-cell interference, $\SumI$, is given by}
\begin{align}\label{eq:Sum_interference}
	\SumI = \sum_{\textbf{x}' \in \Psi_{n'}} \Ptx{n} h_{\textbf{x}',o}d_{\textbf{x}',o}^{-\PathExp},
\end{align}
where $\Psi_{n'}$ is the location of sensors that \blue{use the same frequency resource} with the $n$-th sensor.

The transmission can be successful if the data rate is larger than the target data rate.
When \blue{the Rayleigh fading channel is considered}, i.e., $\ChannelFading{n} \sim \text{exp}(1)$, for the $k$-th data transmission, the outage probability can be given by \cite{TseVis:05}
\begin{align}\label{eq:Outage_prob}
	\begin{aligned}
		\OutageProb{k} &= \mathbb{P}[\Bandwidth \log_{2} (1+\SNR{n}) < R_{\UpdateK}] \\
		&= \mathbb{E}_{\SumI}\left[\mathbb{P}\left[\ChannelFading{n} < \frac{\DistToSink{n}^{\PathExp}(\sigma^2 + \SumI) (2^\frac{R_{\UpdateK}}{\Bandwidth}- 1)}{\Ptx{n}} \right]\right] \\
		&= 1 - \exp \left(- A \sigma^{2}\right) \LaplaceTrans{\SumI}{A},
	\end{aligned}
\end{align}
where $m_k$ indicates the computing decision, i.e., $\UpdateK=\edge$ for \ac{EC} or $\UpdateK=\loc$ for \ac{LC}, $\Bandwidth$ is the bandwidth, and $A = (2^{R_{\UpdateK}/\Bandwidth} - 1) \DistToSink{n}^{\PathExp} / \Ptx{n}$.
Here, $R_{\UpdateK}$ is the target data rate that a sensor needs to transmit within time duration $\SlotDuration$ (i.e., a time slot) \cite{ZhaAdvLim:06}, so $R_{\edge} = \InDataSize/\SlotDuration$ and $R_{\loc} = \OutDataSize/\SlotDuration$.
Using the definition of the Laplace transform, we have \cite[Eq. 3.21]{HaeGan:09}
\begin{align}\label{eq:Laplace_transform}
	\begin{aligned} 
		\LaplaceTrans{\SumI}{A} = \mathbb{E}_{\SumI}\left[e^{- A \SumI}\right] 
		= \exp \left\{-\pi \blue{\Idensity} (A \Ptx{n})^{\frac{2}{\PathExp}} \frac{2 \pi}{\PathExp \sin (2 \pi / \PathExp)} \right\}.
	\end{aligned}
\end{align}
By combining \eqref{eq:Laplace_transform} with \eqref{eq:Outage_prob}, we get the outage probability as
\begin{align}\label{eq:Outage_prob_result}
	\begin{aligned}
		\OutageProb{k} = 1 - \exp \left\{- A \sigma^{2} -\pi \blue{\Idensity} (A \Ptx{n})^{\frac{2}{\PathExp}} \frac{2 \pi}{\PathExp \sin (2 \pi / \PathExp)}  \right\}.
	\end{aligned}
\end{align}
Here, $p_{\textbf{o},{k}} > p_{\textbf{o},{j}}$ where $m_k = \edge$ and $m_j = \loc$ as $R_{\edge} > R_{\loc}$ due to $\InDataSize > \OutDataSize$.

To improve the data freshness at the sink node, we adopt a retransmission scheme so that the sensor can transmit the data up to $\Txcount$ times when the current transmission fails.
If the transmission fails for $\Txcount$ times, the sensor drops the data and becomes idle until it has the new data by sensing to prevent the excessive energy consumption.

\subsection{Energy Model}\label{ssec:Energy_model}
In wireless sensor networks, sensors are equipped with the battery instead of receiving power via the wire since the deployment of sensors for wired sensor networks may be infeasible, especially in hostile area.
Furthermore, the wireless sensor networks have advantages in terms of the installing and maintaining cost, compared to the wired sensor networks \cite{YonBac:05}.

We consider two battery models, generally considered in wireless sensor networks: pre-charged battery model and \ac{EH} battery model.\footnote{\blue{Note that the proposed algorithms can also be applicable for the hybrid energy model, i.e., the battery has both pre-charged energy and harvesting capability, as the same manner, used for the \ac{EH} model case.}}
In this subsection, we describe both models.

\subsubsection{Pre-Charged Battery Model}\label{sssec:Pre_charged_model}
In this model, the battery of the sensor is fully charged in advance \cite{VieCoeSil:03}. 
Hence, when we consider $\TotRound$ rounds (i.e., $\TotRound \RoundLen$ time slots), the consumed energy at the $n$-th sensor is limited as $\sum_{i=1}^{\TotRound \RoundLen} \ConsumeEnergy{n}{i} \leq \PreBatteryTh$, where $\ConsumeEnergy{n}{i} \in \{ \E{\SE}, \E{\T}, \Ep,0 \}$ is the consumed energy of the $n$-th sensor at the $i$-th time slot, and $\PreBatteryTh$ is the battery constraint for $\TotRound$ rounds.

\subsubsection{EH Model}\label{sssec:EH_model}
In this model, sensors can be charged by harvesting energy.
We consider a random energy arrival model with the mean harvested energy $\E{\Harv}$ for each time slot, i.e., \blue{$\mathbb{E}[\HarvEnergy{n}{i}] = \E{\Harv}$, $n \in \SensorSet$, $\forall i$,} where $\HarvEnergy{n}{i}$ is the harvested energy of the $n$-th sensor during the $i$-th time slot \cite{LuoPuWan:19}. 
An evolution of the battery level of the $n$-th sensor is then given by
\begin{align}\label{eq:Energy_flow}
	\EnergyLevel{n}{t} = \min{\left[\EnergyLevel{n}{t_{i}} - \ConsumeEnergy{n}{i} + \HarvEnergy{n}{i}, \Bmax \right]} \geq 0,
\end{align}
for $t_i \leq t < t_{i+1}$, where $t_j$ is the starting time of the $j$-th time slot, and $\Bmax$ is the battery capacity.
Here, $\HarvEnergy{n}{i}$ follows the uniform distribution, i.e., $\HarvEnergy{n}{i} \sim \mathcal{U}[\varepsilon_{\text{min}}, \varepsilon_{\text{max}}]$ where $\varepsilon_{\text{min}}$ and $\varepsilon_{\text{max}}$ are the minimum and maximum values of harvested energy at each time slot, respectively \cite{LeeEuHan:11}.
Note that the sensor can perform sensing, transmission, or \ac{LC} only when the current battery level is sufficient to perform that mode.

\subsection{\Aratio}\label{ssec:Network_coverage}
\begin{figure}[t!]
	\begin{center}   
		{ 
			\includegraphics[width=1\columnwidth]{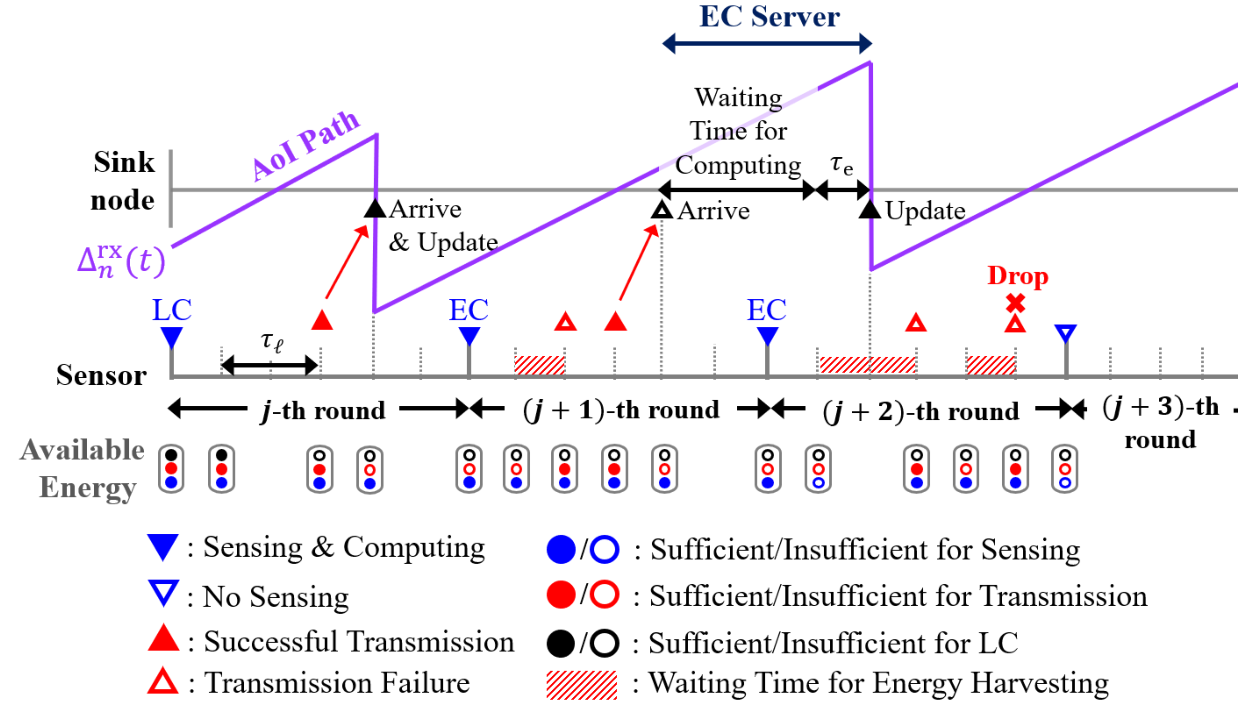}
		}
	\end{center}
	\caption{
		Example path of $\SinkAoI{n}{t}$ for EH model.
	}
	\label{fig:Update_model}
\end{figure}

For the data of the $n$-th sensor, the \ac{AoI} can be defined at both the sensor and the sink node, denoted as $\SensorAoI{n}{t}$ and $\SinkAoI{n}{t}$, respectively.
The \ac{AoI} at the sensor, $\SensorAoI{n}{t}$, becomes $\SlotDuration$ (i.e., one time slot length) when the sensor obtains new data by sensing. Otherwise, the \ac{AoI} increases linearly over the time.
On the other hand, the \ac{AoI} at the sink node, $\SinkAoI{n}{t}$, is defined as the time gap from sensing to the time when the sink node obtains the computed data.
Hence, $\SensorAoI{n}{t}$ and $\SinkAoI{n}{t}$ are respectively given by
\begin{align}
	\SensorAoI{n}{t} = t - \GenTime{n}{t}, \\
	\SinkAoI{n}{t}   = t - \UpdateTime{n}{t}, \label{eq:AoI_eq}
\end{align}
where $\GenTime{n}{t}$ is the latest data generation time at the $n$-th sensor, and $\UpdateTime{n}{t}$ is the generation time of the lately updated data of the $n$-th sensor at the sink node.
\blue{Note that $\SinkAoI{n}{t}$ and $\SensorAoI{n}{t}$ will be used for defining and analyzing the \aratio\ in Section \ref{sec:Network_model}, \ref{sec:Probability_SCD} and the \ac{RL}-based algorithm in Section \ref{sec:MARL_SCD}, respectively.}

Figure \ref{fig:Update_model} represents an example path of $\SinkAoI{n}{t}$ over time for $\RoundLen = 6, \ProcSlotE = 1, \ProcSlotL = 2$, and $\Txcount = 2$ for the \ac{EH} model.
At the beginning of the $j$-th round, the $n$-th sensor decides to perform sensing and choose \ac{LC}. 
After sensing and computing at the sensor, the sensor transmits the computed data to the sink node, and \blue{$\SinkAoI{n}{t}$} is updated once it is successfully received.
At the beginning of the $(j+1)$-th round, the sensor decides to perform sensing and choose \ac{EC}. 
Since the sensor does not have sufficient energy for transmission, it waits to get more energy by harvesting.
Once the data is successfully received at the sink node, \blue{$\SinkAoI{n}{t}$} can be updated after completing the computation.
Here, when the transmission fails $\Txcount$ (e.g., two in this example) times, the data is dropped as shown in the $(j+2)$-th round.

In sensor networks, the sensed data is generally correlated in time and space \cite{Ste:05}.
Hence, we can estimate the data at certain point $\textbf{y}$ at time $t$ with the lately generated or updated data of the $n$-th sensor, located at $\textbf{x}_{n}$, at time $\GenTime{n}{t}$ or $\UpdateTime{n}{t}$.
Here, the estimator of the data that minimizes the mean square error becomes the conditional expectation, which is given by \cite{Ste:05}
\begin{align}\label{eq:Estimator_conditional_expectation}
	\hat{H} = \mathbb{E}\left[H(\textbf{y},t) | H(\textbf{x}_{n}, \WhichGenUpdateTime{n}{t})\right], \WhichGenUpdateTimeIdx \in \{\text{tx, rx}\}, 
\end{align}
where $H(\textbf{z},t)$ is the data at $\textbf{z}$ and the time $t$.

When the sensed data follows a stationary Gaussian process, the correlation coefficient is represented using the covariance model with the \ac{AoI}, i.e., $\WhichAoI{n}{t} = t - \WhichGenUpdateTime{n}{t}$, given by \cite{HriCoKamDa:18}
\begin{align}\label{eq:Covariance}
	\Covariance{n}(\WhichAoI{n}{t}) = \exp{(-\ErrWeight{1} \DistToPoint{n} - \ErrWeight{2} \WhichAoI{n}{t})}, \WhichGenUpdateTimeIdx \in \{\text{tx, rx}\},
\end{align}
where $\DistToPoint{n}$ is the Euclidean distance between $\textbf{x}_{n}$ and $\textbf{y}$, $\ErrWeight{1}$ and $\ErrWeight{2}$ are the weights of the spatial error and temporal error, respectively.
Since the variance of the estimation error of \eqref{eq:Estimator_conditional_expectation} linearly increases with $ 1 - \{\Covariance{n}(\WhichAoI{n}{t})\}^2$ \cite{Ste:05}, the estimation error at $\textbf{y}$ is determined by \eqref{eq:Covariance} as \cite{HriCoKamDa:18}
\begin{align}\label{eq:Estimation_error}
	\begin{aligned}
		\EstimationErr{n}(\WhichAoI{n}{t})
		&= 1 - \{\Covariance{n}(\WhichAoI{n}{t})\}^2 \\
		&= 1 - \exp{(-2\ErrWeight{1} \DistToPoint{n} - 2\ErrWeight{2} \WhichAoI{n}{t})}, \WhichGenUpdateTimeIdx \in \{\text{tx, rx}\}.
	\end{aligned}
\end{align}
From \eqref{eq:Estimation_error}, we can observe that the estimation error increases as $\DistToPoint{n}$ increases or $\WhichAoI{n}{t}$ increases.
We assume the estimated data of $\textbf{y}$ at $t$ from the sensed data at $\textbf{x}_n$ and time $\WhichGenUpdateTime{n}{t}$ becomes invalid when the estimation error, $\EstimationErr{n}(\WhichAoI{n}{t})$, is greater than the threshold $\ErrTh$ as
%
\begin{align}\label{eq:Error_threshold}
	\EstimationErr{n}(\WhichAoI{n}{t}) > \ErrTh, \WhichGenUpdateTimeIdx \in \{\text{tx, rx}\}.
\end{align}
Then, the sensing coverage of the $n$-th sensor can be defined as the area that the sensed data can be used to estimate the data of the area with less error than $\ErrTh$ \cite{KimKimKim:23}.
The sensing coverage becomes a circle with radius $\SensingRad(\WhichAoI{n}{t}, \ErrTh)$, given by
\begin{align}\label{eq:Sensing_radius}
	\SensingRad(\WhichAoI{n}{t}, \ErrTh) = \frac{-2\ErrWeight{2}\WhichAoI{n}{t} - \log(1-\ErrTh)}{2\ErrWeight{1}}, \WhichGenUpdateTimeIdx \in \{\text{tx, rx}\},
\end{align}
which can be obtained by substituting \eqref{eq:Estimation_error} into \eqref{eq:Error_threshold} for $\EstimationErr{n}(\WhichAoI{n}{t}) = \epsilon_{0}$.
Since the sink node collects and manages the information of sensors, we focus on the \ac{AoI} at the sink node, i.e., $\SinkAoI{n}{t}$, and from \eqref{eq:Sensing_radius}, we can see that $\SensingRad(\SinkAoI{n}{t}, \ErrTh)$ decreases as $\SinkAoI{n}{t}$ increases.

Next, we define the error-tolerable network coverage ratio from the sensing coverage model.
We consider a 2-D discrete grid network $\Coordinate$.
We define an event $\PointSensorCov{n}(t)$ that the point at $\textbf{y}$ is in the coverage of the $n$-th sensor at time $t$, given by
\blue{
	\begin{align}\label{eq:Coverage_for_sensor}
		\PointSensorCov{n}(t) = 
		\begin{cases}
			1, &\DistToPoint{n} \leq \SensingRad(\SinkAoI{n}{t}, \ErrTh), \\
			0, & \mbox{otherwise}.
		\end{cases}
	\end{align}
}
Since the sensor network consists of multiple sensors, we can regard the point at $\textbf{y}$ is in the network coverage if it is covered by at least one sensor.
Here, the sink node uses the data which has the minimum estimation error among the sensors, i.e., $\displaystyle \min_{n \in \SensorSet} \EstimationErr{n}(\SinkAoI{n}{t}) $.
Then, whether the point at $\textbf{y}$ is covered or not in the network \blue{is} indicated as $\PointCov \in \{0,1\}$, given by
\begin{align}\label{eq:Coverage_point}
	\PointCov(t) = 1 - \hspace*{0mm} \prod_{n\in \SensorSet} \hspace*{0mm} (1 - \PointSensorCov{n}(t)) .
\end{align}
From \eqref{eq:Coverage_point}, at time $t$, we can define the network coverage ratio $\NetCov(t)$ as the ratio of the area, covered by at least one sensor, to the whole network area $|\Coordinate|$ as
\begin{align}\label{eq:Coverage_rate}
	\NetCov(t) = 
	\frac{\sum_{\textbf{y} \in \Coordinate} \PointCov(t)} 
	{|\Coordinate|}.
\end{align}
%
Finally, we define the \aratio\ $\ECovProb$ as the probability that $\NetCov(t)$ is larger than a target coverage ratio $\TgtCov$, which is given by \cite{KimKimKim:23}
\begin{align}\label{eq:Aratio}
	\ECovProb =  \mathbb{P}[\NetCov(t) \geq \TgtCov].
\end{align}
Note that $\ECovProb$ can also show the portion of the network area where the sink node has the data with smaller error than $\ErrTh$.

\subsection{Two SCD Approaches}\label{ssec:two_SCD_approaches}
To enhance the \aratio, we can design the \ac{SCD} algorithm in two approaches: \blue{1)} the probability-based \ac{SCD} and \blue{2)} the \ac{RL}-based \ac{SCD}.
\blue{For the probability-based \ac{SCD}, the optimal probabilities for sensing and \ac{EC}, which maximize the \aratio\ in average sense, are used without requiring the real-time information from other sensors or \ac{EC} server.
	However, in this algorithm, the sensors cannot utilize network status such as the \ac{AoI} and battery level, which may lead to a low performance.
	On the other hand, in the \ac{RL}-based \ac{SCD}, each sensor makes the decision based on the current network status to improve the \aratio\ through the training process.
	Specifically, the real-time information of the network status such as the \ac{AoI} and the battery level are needed, but it generally achieves the higher performance than the probability-based \ac{SCD} algorithm as it makes decisions based on the current status information.}\footnote{\blue{Note that the probability-based \ac{SCD} algorithm has an advantage in terms of implementation simplicity, and the \ac{RL}-based \ac{SCD} algorithm can perform worse when the status information from neighboring sensors and the \ac{EC} server are not reliably provided.}} 
In the following sections, we provide the probability-based \ac{SCD} algorithm (Section \ref{sec:Probability_SCD}) and the \ac{RL}-based \ac{SCD} algorithm (Section \ref{sec:MARL_SCD}).

\section{Probability-based SCD Algorithm}\label{sec:Probability_SCD}
In this section, we propose the probability-based \ac{SCD} algorithm, which determines to sense with the probability $\Ps$ and determines to compute at the \ac{EC} server with the probability $\Pe$ at each round.
To maximize the \aratio, we optimize $\Ps$ and $\Pe$ in the probability-based \ac{SCD} algorithm.
For that, in this section, we first derive the \aratio\ for the single pre-charged sensor case, and provide the optimal $\Ps$ and $\Pe$.
We then discuss how to obtain the optimal $\Ps$ and $\Pe$ for \blue{the multiple} \ac{EH} sensor case.
\begin{figure}[t!]
	\begin{center}   
		{ 
			\includegraphics[width=1\columnwidth]{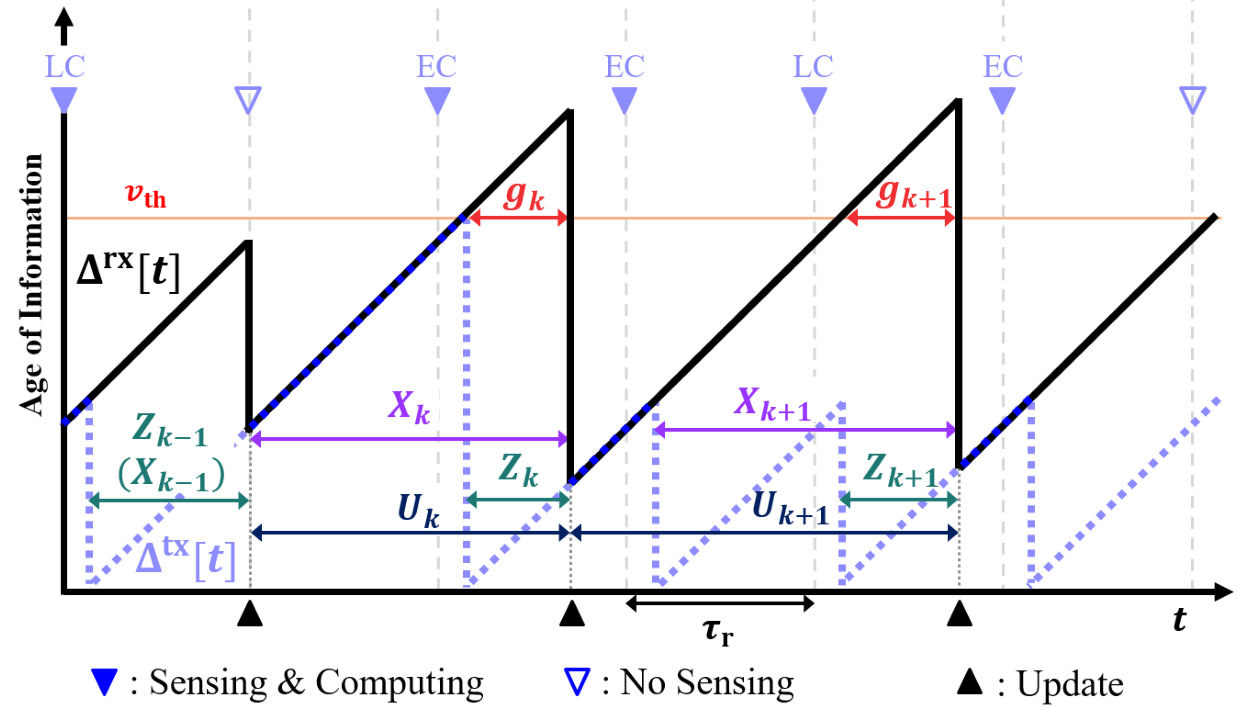}
		}
	\end{center}
	\caption{
		Example of the \ac{AoI} path of the single pre-charged sensor case, $\SensorAoISgl{t}$ and $\SinkAoISgl{t}$, from the $(k-1)$-th to the $(k+1)$-th successful update.
	}
	\label{fig:AoI_single}
\end{figure}

\subsection{Single Pre-Charged Sensor Case}\label{ssec:prob_SCD_single}
\subsubsection{\Aratio\ Analysis}\label{sssec:prob_SCD_single_analysis}
We first analyze the \aratio\ in the single pre-charged sensor case.
Here, we consider a 2-D network $\CoordinateSingle$ with a sink node and a sensor, and omit index $n$ for simplicity.
In this case, $\PointCov(t)$ in \eqref{eq:Coverage_point} becomes $c_{\textbf{x},\textbf{y}}(t)$, and in \eqref{eq:Coverage_rate}, we have $\sum_{\textbf{y} \in \CoordinateSingle} \PointCov(t) = \sum_{\textbf{y} \in \CoordinateSingle}c_{\textbf{x},\textbf{y}}(t)$, which indicates the sensing coverage of the sensor.
The radius of the sensing coverage is $\SensingRad(\SinkAoISgl{t}, \ErrTh)$, so the sensing coverage area can be presented as $\pi \SensingRad^2(\SinkAoISgl{t}, \ErrTh)$.\footnote{When the resolution of grid is high in the discrete grid networks, the sensing coverage area can be approximated as the circle with subtle error.}
When the sensing coverage is completely included in $\CoordinateSingle$, $\NetCov(t)$ can be given by
%
\begin{align}\label{eq:Single_coverage}
	\NetCov(t) = \frac{\pi \SensingRad^2(\SinkAoISgl{t}, \ErrTh)}{|\CoordinateSingle|} =
	\frac{\pi \{-2\ErrWeight{2} \SinkAoISgl{t} - \log(1-\ErrTh)\}^2}{4\ErrWeight{1}^2|\CoordinateSingle|},
\end{align}
where $|\CoordinateSingle|$ is the considered network area in the single sensor case.
From \eqref{eq:Aratio} and \eqref{eq:Single_coverage}, $\ECovProb$ is given by  
\begin{align}\label{eq:Aratio_AVP}
	\begin{aligned}
		\ECovProb
		&= \mathbb{P}\left[\frac{\pi \{-2\ErrWeight{2} \SinkAoISgl{t} - \log(1-\ErrTh)\}^2}{4\ErrWeight{1}^2|\CoordinateSingle|} \geq \TgtCov\right] \\
		&= 1 - \mathbb{P}\left[\SinkAoISgl{t} > \vth \right],
	\end{aligned}  
\end{align}
where the target \ac{AoI} is $\vth = -\frac{\ErrWeight{1}}{\ErrWeight{2}} \sqrt{\frac{\TgtCov |\CoordinateSingle|}{\pi}} - \frac{\log(1-\ErrTh)}{2\ErrWeight{2}}$. From \eqref{eq:Aratio_AVP}, we can see that in the single pre-charged sensor case, the \aratio\ can be represented as the \AVP\ $\mathbb{P}\left[\SinkAoISgl{t} > \vth \right]$, which is the probability that the \ac{AoI}, $\SinkAoISgl{t}$, violates the target \ac{AoI} $\vth$.

For the analysis of the \aratio, we first define some time intervals in the \ac{AoI} path as follows.
An example of the \ac{AoI} path of the single pre-charged sensor case, $\SensorAoISgl{t}$ and $\SinkAoISgl{t}$, from the $(k-1)$-th to the $(k+1)$-th successful update are presented in Fig. \ref{fig:AoI_single}.
For the $k$-th successful update, we denote the time interval from the sensing instance to the successful update as $Z_k$, the time interval from the beginning of the first round after the $(k-1)$-th successful update to the $k$-th successful update time as $X_k$, the inter update time as $U_k$, and the violation time for $k$-th successful update as $g_k$.
We use $\UpdateK \in \{\text{e}, \loc\}$ to indicate whether the information of the $k$-th successful update is computed by the \ac{EC} ($\UpdateK=\text{e}$) or \ac{LC} ($\UpdateK=\loc$).

Next, we analytically obtain the \aratio\ in \eqref{eq:Aratio_AVP} for the single pre-charged sensor case.
we first assume sensing, computing, and transmission caused by the current round are completed before starting the next round, i.e., $1 + \Txcount + \tk \leq \RoundLen, \tk \in \{\ProcSlotE, \ProcSlotL\}$.
Note that for the single sensor case, we assume that the waiting time at the \ac{EC} server is negligible as it only handles the information from one sensor.

The \AVP\ is given by \cite{ChaAlGro:19}
\begin{align}\label{eq:AVP}
	\mathbb{P} \left[ \SinkAoISgl{t} > \vth \right] = \frac{\gk}{\Uk},
\end{align}
%
where $\Uk$ and $\gk$ are the expected inter-update and violation time, respectively.
Here, $U_k$ and $g_k$ \blue{are presented} as
%
\begin{align}\label{eq:Inter_update_time}
U_k = \RoundLen - Z_{k-1} + X_k,
\end{align}
\begin{align}\label{eq:gk_def}
g_k = 
\begin{cases}
	\RoundLen + X_k - Z_{k-1}, &\mathrm{if}\quad \AVPCaseOne,\\
	\RoundLen + X_k - (\vfloor+1), &\mathrm{if}\quad \AVPCaseTwo,\\
	0, &\mathrm{if}\quad \AVPCaseThree,
\end{cases}
\end{align}
where $\lfloor x \rfloor$ is a floor function that gives the greatest integer less than or equal to $x$.
In the following lemma, we obtain $\Uk$ for given $\Ps$ and $\Pe$.

\begin{lemma}[Expected Inter-Update Time]
The expected inter-update time $\Uk$ in \eqref{eq:AVP} is obtained as
\begin{align}\label{eq:Uk_result}
	\Uk = \frac{\RoundLen}{\Px},
\end{align}
where $\Px$ is given by
\begin{align}\label{eq:Px}
	\begin{aligned}
		\Px &= \Ps [\Pe \{\TxProbPout{\text{e}}{\Txcount}\}+ (1-\Pe)\{\TxProbPout{\loc}{\Txcount}\}].
	\end{aligned}
\end{align}
\end{lemma}
\begin{IEEEproof}
See Appendix \ref{app:Lemma_1}.
\end{IEEEproof}

In the following lemma, we also obtain $\gk$ for given $\Ps$ and $\Pe$.	

\begin{lemma}[Expected Violation Time]
$\gk$ in \eqref{eq:AVP} is obtained as
\begin{align}\label{eq:E_gk_def}
	\begin{aligned}
		\gk = \sum_{\UpdateKMO \in \{\text{e}, \loc \}} \sum_{\UpdateK \in \{\text{e}, \loc \}}
		\Peprime{\UpdateKMO}\Peprime{\UpdateK} G(\UpdateKMO,\UpdateK),
	\end{aligned}
\end{align}
where $\Peprime{\UpdateK}$ is the probability that \ac{EC} ($\UpdateK=\text{e}$) or \ac{LC} ($\UpdateK=\loc$) is chosen, conditioned on the successful update, given by
\begin{align}\label{eq:cndt_edge_prob}
	\Peprime{\UpdateK} = 
	\begin{cases}
		\frac {\Pe \{\TxProbPout{\text{e}}{\Txcount}\}}
		{\Pe \{\TxProbPout{\text{e}}{\Txcount}\}+ (1-\Pe)\{\TxProbPout{\loc}{\Txcount}\}}, &\mathrm{if}\quad \UpdateK = \text{e},\\
		\frac {(1-\Pe)\{\TxProbPout{\loc}{\Txcount}\}}
		{\Pe \{\TxProbPout{\text{e}}{\Txcount}\}+ (1-\Pe)\{\TxProbPout{\loc}{\Txcount}\}}, &\mathrm{if}\quad \UpdateK = \loc.
	\end{cases}
\end{align}
Here, $G(\UpdateKMO,\UpdateK)$ is presented in \eqref{eq:gk_cndt_result_onecol} according to the range of $\vth$, where
\begin{align} \label{eq:vth_range_def}
	\begin{aligned}
		v_{1,k} &= 2+\tkmo, \\
		v_{2,k} &= 1 + \Txcount + \tkmo, \\
		v_{3,k}(\y) &= \y \RoundLen + 2 + \tk,\\
		v_{4,k}(\y) &= \y \RoundLen + 1 +\Txcount+\tk,\\
	\end{aligned}
\end{align}  
for $\y=1,2,\cdots$. In \eqref{eq:gk_cndt_result_onecol}, $\TxProbxTime{k}{c}, \FuncOne{k}{\Txcount}, \FuncTwo{k}{\y}{\Txcount}$ and $\FuncThree{k}{\y}$ are respectively defined as
\begin{align}
	\TxProbxTime{k}{c} &= \TxProbPout{k}{c}, c \in \{1,2,\cdots, \Txcount\}, \label{eq:Tx_prob_time} \\
	\FuncOne{k}{\Txcount} &= \frac{1}{\TxProbxTime{k}{1}} -
	\frac{\Txcount \OutageProb{k}^{\Txcount} \TxProbxTime{k}{1}}{\TxProbPf{k}},\label{eq:FuncOne} \\
	\FuncTwo{k}{\y}{\Txcount} &= 1 + \Txcount + \tk + \RoundLen(\y-1), \label{eq:FuncTwo}\\
	\FuncThree{k}{\y} &= \vfloor - \tk - \RoundLen \y - 1. \label{eq:FuncThree}
\end{align}
\end{lemma}
\begin{figure*}[t!]
\normalsize
$G(\UpdateKMO,\UpdateK) =$ 
\begin{align}
	\begin{aligned}\label{eq:gk_cndt_result_onecol}
		\begin{cases}
			\frac{\RoundLen}{\Px} + \tk - \tkmo + \FuncOne{k}{\Txcount}
			- \FuncOne{k-1}{\Txcount}, 
			& \vth < v_{1,k},\\
			%
			\frac{\RoundLen}{\Px} + \tk - \tkmo + \FuncOne{k}{\Txcount}
			- \frac{\vfloor-\tkmo-\OutageProb{k-1}^{\vfloor -1 - \tkmo} - \Txcount \OutageProb{k-1}^{\Txcount}}
			{\TxProbPf{k-1}}
			- \frac{\OutageProb{k-1}^{\vfloor -1 - \tkmo} - \OutageProb{k-1}^{\Txcount}}
			{\TxProbPf{k-1}\TxProbxTime{k-1}{1}},
			& v_{1,k} \leq \vth < v_{2,k},\\
			%
			\frac{\RoundLen}{\Px} + \tk +\FuncOne{k}{\Txcount} - \vfloor, 
			& v_{2,k} \leq \vth < v_{3,k}(\y=1), \\
			%
			(1-\Px)^{\y-1}
			\left[(1-\Px)
			\left\{ 
			\FuncTwo{k}{\y}{\Txcount}
			+ \frac{\RoundLen}{\Px}
			- \frac{\Txcount}{\TxProbPf{k}}
			+ \frac{1}{\TxProbxTime{k}{1}}
			\right\} 
			+ (1 - \RoundLen + \vfloor) 
			\left\{
			\frac{\Px \TxProbxTime{k}{\FuncThree{k}{\y}}}{\TxProbPf{k}} - 1
			\right\}
			\right. \\
			\left.
			\qquad\qquad\qquad	+ \frac{\Px}{\TxProbPf{k}} 
			\left\{
			\TxProbxTime{k}{1}^{\FuncThree{k}{\y}}(\vfloor-\RoundLen) 
			- \TxProbxTime{k}{1}^{\Txcount} \FuncTwo{k}{\y}{\Txcount}
			+ \frac{\TxProbxTime{k}{1}^{\FuncThree{k}{\y}}- \TxProbxTime{k}{1}^{\Txcount}}{\OutageProb{k}}
			\right\}   			
			\right], 
			& v_{3,k}(\y) \leq \vth < v_{4,k}(\y), \\
			%
			(1- \Px)^\y 
			\left\{
			- \FuncThree{k}{\y}-1
			+\frac{\RoundLen}{\Px}
			+\frac{1}{\TxProbxTime{k}{1}}
			+\Txcount\left(1 - \frac{1}{\TxProbPf{k}}\right)
			\right\}, 
			& v_{4,k}(\y) \leq \vth < v_{3,k}(\y+1). \\
		\end{cases}
	\end{aligned}
\end{align}
\centering \rule[0pt]{18cm}{0.3pt}
\end{figure*}
\begin{IEEEproof}
See Appendix \ref{app:Lemma_2}. 
\end{IEEEproof}

From Lemma 1 and Lemma 2, the \AVP\ $\mathbb{P} \left[ \SinkAoISgl{t} > \vth \right]$ is then derived by dividing $\gk$ into $\Uk$ as \eqref{eq:AVP}.
Finally, by using \eqref{eq:AVP} in \eqref{eq:Aratio_AVP}, we can obtain the \aratio\ for the single pre-charged sensor case for given $\Ps$ and $\Pe$, obtained as
\begin{align}\label{eq:aratio_result}
\begin{aligned}
	&\ECovProbPsPe \\
	&= 1-\frac{\Px}{\RoundLen}\sum_{\UpdateKMO \in \{\text{e}, \loc \}} \sum_{\UpdateK \in \{\text{e}, \loc \}}
	\Peprime{\UpdateKMO}\Peprime{\UpdateK} G(\UpdateKMO,\UpdateK).
\end{aligned}
\end{align}
Here, we change the notation $\ECovProb$ to $\ECovProbPsPe$ to use it for optimizing $\Ps$ and $\Pe$ in the following section.

\subsubsection{Optimal Probability of Sensing and Computing}\label{sssec:Prob_SCD_single_opt}
We obtain the optimal $\Ps$ and $\Pe$ that maximize the \aratio.
For that, we first obtain the average energy consumption per the time duration of a round $\RoundLen$, denoted as $\bar{E}_{t}$.
The average energy consumption over $\RoundLen$ when the sensor chooses \ac{EC}, $\bar{E}_\text{e}$, is given by
\begin{align}\label{eq:Energy_edge}
\begin{aligned}
	\bar{E}_\text{e} 
	= \Es + \Et \AvgTxNum{\text{e}}{\text{e}},
\end{aligned}
\end{align}
where $\AvgTxNum{\UpdateK}{k}$ is the average number of transmissions, given by 
\begin{align} \label{eq:Average_Tx_Num}
\AvgTxNum{\UpdateK}{k}= \sum_{c=1}^{\Txcount} c \OutageProb{k}^{c-1}(1-\OutageProb{k}) + \Txcount \OutageProb{k}^{\Txcount} = \frac{\TxProbPf{k}}{\TxProbxTime{k}{1}}.
\end{align}
Similarly, the average energy consumption over $\RoundLen$ when the sensor chooses \ac{LC}, $\bar{E}_\loc$, is given by
\begin{align}\label{eq:Energy_local}
\begin{aligned}
	\bar{E}_\loc 
	&= \Es + \Ep + \Et \AvgTxNum{\loc}{\loc}.
\end{aligned}
\end{align}
For given $\Ps$ and $\Pe$, $\bar{E}_{t}$ is then obtained as
\begin{align}\label{eq:Energy_const_prob_singlepre}
\begin{aligned}
	\bar{E}_{t} = \Ps\Pe \bar{E}_\text{e} + \Ps(1 - \Pe) \bar{E}_\loc.
\end{aligned}
\end{align}
Note that \blue{the consumed energy over $\TotRound$ rounds} needs to be less than or equal to $\PreBatteryTh$ in the pre-charged battery model, i.e., $\TotRound \bar{E}_{t} \leq \PreBatteryTh$, as described in Section \ref{ssec:Energy_model}.

We now formulate the \aratio\ maximization problem for the single pre-charged sensor case.

\textit{Problem 1 (\Aratio\ Maximization in Single Pre-Charged Sensor Case):}
\begin{align}
\max_{\Ps,\Pe} \qquad                                                                                                                                                                                                                                                                                                                                     
&\ECovProbPsPe \\
\text{s.t.} \qquad 
& \blue{\TotRound \bar{E}_{t} \leq \PreBatteryTh,} \\
& 0 \leq \Ps \leq 1, \label{eq:ps_const} \\
& 0 \leq \Pe \leq 1. \label{eq:pe_const} 
\end{align}
To solve Problem 1, we first discuss the effect of $\Ps$ on $\ECovProbPsPe$ in the following remark.
%
\begin{remark}
For given $\Pe$, $\ECovProbPsPe$ is a monotonically increasing function of $\Ps$. 
For example, for $v_{1,k} \leq \vth < v_{2,k}$ for both $\UpdateKMO=\text{e}$ and $\UpdateKMO=\loc$, i.e., $2+\ProcSlotE \leq 2+\ProcSlotL \leq \vth < 1 + \Txcount + \ProcSlotE \leq  1 + \Txcount + \ProcSlotL$, the first derivative of $\ECovProbPsPe$ with respect to $\Ps$ is given by
\begin{align} \label{eq:Derivate_ps_Case2}
	\begin{aligned}
		\frac{\partial \ECovProbPsPe}{\partial \Ps} =
		\frac{1}{\RoundLen}
		&\Bigg[ \vfloor - \{\Pe\ProcSlotE + (1-\Pe)\ProcSlotL \}\\ 
		&- \left\{\Pe\frac{\TxProbxTime{\text{e}}{\vfloor-\ProcSlotE}}{\TxProbxTime{\text{e}}{1}} 
		+ (1-\Pe) \frac{\TxProbxTime{\loc}{\vfloor-\ProcSlotL}}{\TxProbxTime{\loc}{1}} \right\}
		\Bigg].
	\end{aligned}
\end{align}
Here, from \eqref{eq:Tx_prob_time}, $\TxProbxTime{\text{e}}{\vfloor-\ProcSlotE} / \TxProbxTime{\text{e}}{1}$ and $\TxProbxTime{\loc}{\vfloor-\ProcSlotL} / \TxProbxTime{\loc}{1}$ in \eqref{eq:Derivate_ps_Case2} are expressed as
\begin{align} \label{eq:Tx_prob_frac}
	\begin{aligned}
		\frac{\TxProbxTime{k}{\vfloor-\tau_k}}{\TxProbxTime{k}{1}} 
		& =\frac{\TxProbPout{k}{\vfloor-\tau_k}}{1-\OutageProb{k}} \\
		&= \sum_{c=0}^{\vfloor-\tau_k -1} (\OutageProb{k})^{c} < \vfloor-\tau_k, \UpdateK \in \{\text{e},\loc\},
	\end{aligned}
\end{align}
where the inequality in \eqref{eq:Tx_prob_frac} holds since the outage probability is less than 1, i.e., $\OutageProb{k} < 1$.
Then, from \eqref{eq:Tx_prob_frac}, \eqref{eq:Derivate_ps_Case2} can be obtained as
\begin{align} \label{eq:Derivate_ps_Case2_inequality}
	\begin{aligned}
		\frac{\partial \ECovProbPsPe}{\partial \Ps} &>
		\frac{1}{\RoundLen}
		\Big[ \vfloor - \{\Pe\ProcSlotE + (1-\Pe)\ProcSlotL \} \\
		&- \left\{\Pe (\vfloor-\ProcSlotE) + (1-\Pe)(\vfloor-\ProcSlotL)\right\} \Big] 
		= 0.
	\end{aligned}
\end{align}
Thus, $\ECovProbPsPe$ is a monotonically increasing function of $\Ps$.\footnote{Similarly, by differentiating  $\ECovProbPsPe$ with respect to $\Ps$, it is also proved that $\ECovProbPsPe$ is a monotonically increasing function of $\Ps$ for other cases.}
\end{remark}

Let $\Ps^{\star}(\Pe)$ denote an optimal sensing probability for given $\Pe$.
We equivalently convert the constraint in \eqref{eq:Energy_const_prob_singlepre} to
\begin{align}\label{eq:Energy_const_ps}
\Ps \leq \frac{\PreBatteryTh}{\TotRound \{\bar{E}_\text{e} \Pe + \bar{E}_\loc (1 - \Pe)\}}.
\end{align}
%
Therefore, from \eqref{eq:Energy_const_ps} and Remark 1, $\Ps^{\star}(\Pe)$, is obtained as
\begin{align}\label{eq:Ps_opt_for_pe}
\Ps^{\star}(\Pe) = 	
\min{\left[
	\frac{\PreBatteryTh}
	{\TotRound \{\bar{E}_\text{e} \Pe + \bar{E}_\loc (1 - \Pe)\}},
	1\right]}.
\end{align}
Unlike $\Ps$, it is not guaranteed that $\ECovProbPsPe$ monotonically increases or decreases with $\Pe$.
Specifically, the trend of $\ECovProbPsPe$ with $\Pe$ depends on parameters such as the outage probability, the computing energy, and the computing time.
Therefore, we use an exhaustive search to obtain $(\Ps^*, \Pe^*)$, i.e., $\Pe^{*}= \argmax_{\Pe} \mathcal{P}^{\TgtCov}_{\text{c}}\Big(\Ps^{\star}(\Pe),\Pe\Big)$ and $\Ps^* = \Ps^{\star}(\Pe^*)$.

\subsection{Multiple EH Sensor Case}\label{ssec:prob_SCD_multiple}
In the multiple \ac{EH} sensor case, it is difficult to clearly analyze the effect of $\Ps$ and $\Pe$ on $\ECovProbPsPe$ by deriving the \AVP\ due to the overlapped coverage of the sensors.
Furthermore, it is hard to formulate the optimization problem for the probability-based \ac{SCD} algorithm \blue{in {EH} case since the energy level of sensor dynamically changes 
and the sensor cannot perform sensing even it is decided to sense when the battery is insufficient.
Thus, in the multiple \ac{EH} sensor case, we need to obtain the optimal solution $(\Ps^{*},\Pe^{*})$ using the exhaustive search by evaluating the \aratio\ using simulation.}\footnote{\blue{Note that for sparse sensor networks, i.e., when sensors rarely have the overlapping area in their coverage, 
the optimal probabilities of single pre-charged sensor (obtained in Section \ref{ssec:prob_SCD_single}) can be used as a guideline for the ones of the multiple \ac{EH} sensor case.}}

In the probability-based \ac{SCD} algorithm, even though $(\Ps^*,\Pe^*)$ can be obtainable, it is difficult to obtain the high \aratio\ since this algorithm cannot consider the real-time dynamics.
For instance, when the battery level is low, decreasing the frequency of sensing may increase the \aratio\ by reducing the energy consumption. 
However, the sensor in the probability-based \ac{SCD} algorithm cannot make dynamic decision.
Therefore, in the following section, we develop an algorithm where each sensor can make real-time decision in the dynamic and complex environment to enhance the \aratio\ in the \ac{EC}-enabled wireless sensor networks.
%

\section{RL-based SCD Algorithm}\label{sec:MARL_SCD}
In this section, we propose the \ac{RL}-based \ac{SCD} algorithm to maximize the \aratio.
Firstly, we formulate a \ac{POMDP} problem for \blue{both the single pre-charged and the multiple \ac{EH} sensor cases}, and then develop an \ac{RL} algorithm and a network architecture to solve the \ac{POMDP} problems. 

\subsection{POMDP Formulation} \label{ssec:MARL_problem_formulation}
In the \ac{RL}-based \ac{SCD} algorithm, sensors are trained to make the decision in the way of maximizing the \aratio.
In this algorithm, we consider an episodic environment with the episode length $T_{\text{ep}}$, where each episode consists of $\TotRound$ rounds, i.e., $T_{\text{ep}} = \TotRound \RoundLen$.

Next, we consider an $\SensorNum$-agent \ac{POMDP}, where the $n$-th agent makes an action $a_{n}[\RoundIdx]$ based on the observation $o_{n}[\RoundIdx]$ and obtains a reward $r_{n}[\RoundIdx]$ at the $\RoundIdx$-th round.
Each agent aims to maximize a return $R_{n}=\sum_{\RoundIdx=1}^{\TotRound} \gamma^{\RoundIdx-1} r_{n}[\RoundIdx]$, where $\gamma\in[0,1]$ is a discount factor.
In the presented \ac{EC}-enabled wireless sensor networks, the \aratio\ maximization problem
can be formulated as the \ac{POMDP}, where each sensor acts as an agent. The observation, action, and reward of each sensor are given as follows.
\begin{itemize}
\item \textbf{Observation}: Each sensor has its observation range $\ObsRange$. 
Then, the observed sensor set at the $n$-th sensor, $\tilde{\SensorSet}_{n}$, is given by
\begin{align}\label{eq:neighbor_set}
	\tilde{\SensorSet}_{n} = \{m|d_{\textbf{x}_{n},\textbf{x}_{m}} \leq \ObsRange, m \in \SensorSet\},
\end{align}
where $d_{\textbf{x}_{n},\textbf{x}_{m}}$ is the distance between the $n$-th and $m$-th sensors. 
When the $\RoundIdx$-th round begins, i.e., $\RoundIdxTime{\RoundIdx}$, the observation of the $n$-th sensor contains the information of sensors in $\tilde{\SensorSet}_{n}$, given by
\begin{align}\label{eq:Observation}
	o_{n}[\RoundIdx] = (\{\EnergyLevel{m}{\RoundIdxTime{\RoundIdx}},\SinkAoI{m}{\RoundIdxTime{\RoundIdx}}, \SensorAoI{m}{\RoundIdxTime{\RoundIdx}}|m \in \tilde{\SensorSet}_{n} \},
	s_{n}^{\text{w}}(\RoundIdxTime{\RoundIdx})).
\end{align}
As shown in \eqref{eq:Observation}, the observation contains the information about the battery level, the \ac{AoI} at the sensor and the sink node side of sensors in $\tilde{\SensorSet}_{n}$, and the local information on the waiting time at the \ac{EC} server, $s_{n}^{\text{w}}(\RoundIdxTime{\RoundIdx})$.

\item \textbf{Action}: Each sensor independently makes \ac{SCD} as an action at the start of each round. 
The action for the $n$-th sensor at the $\RoundIdx$-th round is denoted as $a_{n}[\RoundIdx] \in \{\text{EC, LC, IDLE}\}$, 
where $\text{EC}$ denotes the sensing with \ac{EC}, $\text{LC}$ denotes \blue{the sensing} with \ac{LC}, and $\text{IDLE}$ denotes no sensing and computing.
Here, the sensor can choose $\text{EC}$ or $\text{LC}$ only when the battery level of the sensor is sufficient for performing sensing, i.e.,  $\EnergyLevel{n}{\RoundIdxTime{\RoundIdx}} \geq \Es$. 

\item \textbf{Reward}: The objective of the problem is to maximize the \aratio. For this, in the episodic environment, we set the reward of the $n$-th sensor as
\begin{align}\label{eq:reward}
	r_n [\RoundIdx] = \sum_{i=0}^{\RoundLen-1} 
	\left(\mathbbm{1}[\NetCov(t') \geq \TgtCov] - \RewardPenalty \mathbbm{1}[\NetCov(t') < \TgtCov]\right),
\end{align}
where $t'=\{(\RoundIdx - 1) \RoundLen + i\} \SlotDuration$ and $\RewardPenalty$ is the penalty factor.
Here, we set all sensors to get the shared reward, i.e.,  $r_{n}[\RoundIdx] = r[\RoundIdx], \forall n \in \SensorSet$.
\end{itemize}
\begin{figure}[t!]
\begin{center}   
	{ 
		\includegraphics[width=1\columnwidth]{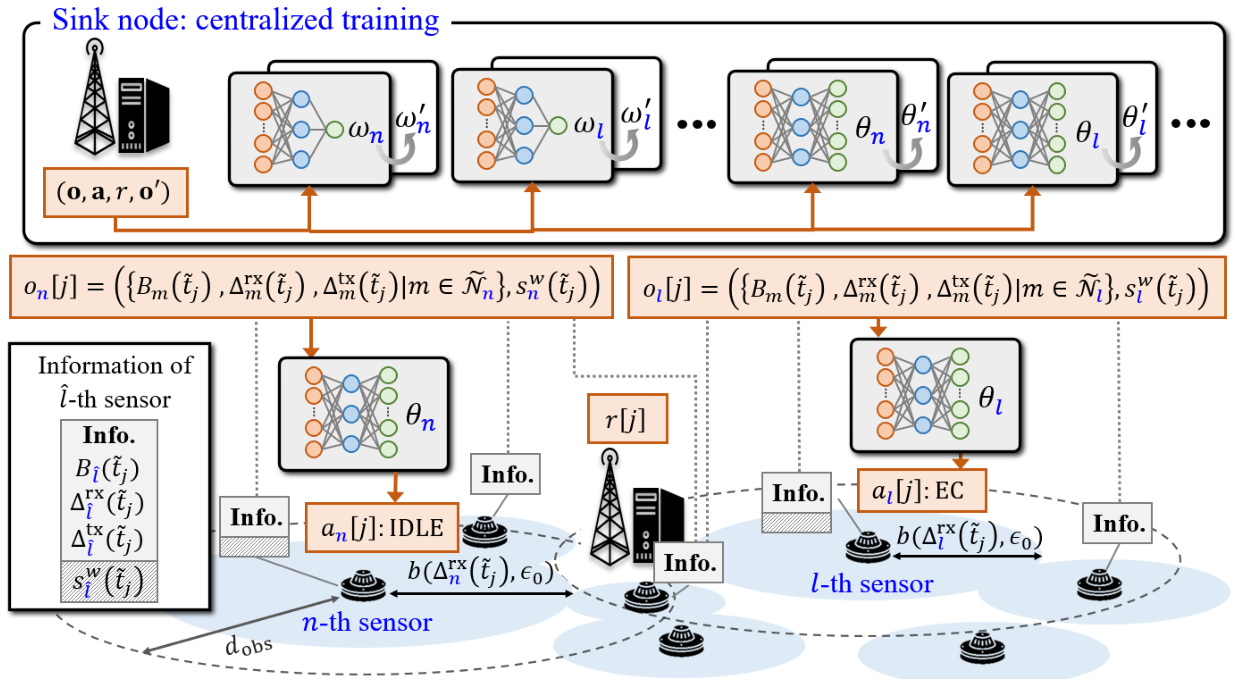}
	}
\end{center}
\caption{
	Proposed \ac{RL}-based \ac{SCD} algorithm. For readability, the neural networks for only two sensors are shown in the figure.
}
\label{fig:Proposed_algorithm}
\end{figure}

\subsection{Proposed Network Architecture}\label{ssec:RL_SCD_EH}
We develop the \ac{RL}-based \ac{SCD} algorithm by implementing the \ac{CTDE} framework, as shown in Fig. \ref{fig:Proposed_algorithm}.
The proposed algorithm contains the actor and critic networks for each sensor. 
Moreover, the target actor and the critic networks are separated from the original networks to stabilize the training.
Specifically, each sensor makes the decision using its actor network, which can reduce the action dimension and communication burden compared to the centralized execution framework.
On the other hand, the critic network estimates the action-value function and aims to minimize the temporal-difference error.
In addition, the sink node collects the observation, the action, and the reward from the sensors to centrally train the actor and critic network parameters.

The pseudo-code of the proposed algorithm is given in \textbf{Algorithm}.
The objective of the proposed algorithm is to train the actor network and critic network of the sensors, parameterized by $\theta_{n}$ and $\omega_{n}$, respectively.
Moreover, the target actor and the critic networks are separated from the original networks to stabilize the learning, parameterized by $\theta'_{n}$ and $\omega'_{n}$, respectively.
We train the neural networks over the $T_{\text{train}}$ episodes, and at the beginning of each episode, observations, actions, and rewards of the sensors are initialized (Line 2).
Here, the bold symbols $\mathbf{o}$ and $\mathbf{a}$ are the observations and actions of the all sensors in the current round, i.e., $\mathbf{o} = \{o_{1},\cdots,o_{\SensorNum}\}$ and $\mathbf{a} = \{a_{1},\cdots,a_{\SensorNum}\}$, respectively.\footnote{For simplicity, the round index $j$ is omitted in this subsection.}

\begin{algorithm}[t!]
\caption{Proposed RL-based SCD algorithm.} \label{algo:Algorithm}
\begin{algorithmic}[1]
	\Statex \textbf{Output}: trained network parameters $\omega_{n}, \theta_{n}, n \in \SensorSet$ 
	\For {episode $= 1 : T_{\text{train}}$}
	\State Initialize observation $\mathbf{o}$, action $\mathbf{a}$, reward $r$
	\For {round $j = 1 :  \TotRound$}
	\For{$n=1 : \SensorNum$}
	\State Select action $a_{n}= G_{\text{gumbel}}(\mu_{\theta_{n}}(o_{n}))$ 
	\EndFor
	\For {$t=\RoundIdxTime{\RoundIdx} : \SlotDuration : \RoundIdxTime{\RoundIdx+1}$}
	\State Update the network status and calculate $\NetCov(t)$
	\EndFor
	\State Get shared reward $r$, new observation $\mathbf{o}'$
	\State Store $(\mathbf{o,a},r,\mathbf{o}')$ in replay buffer
	\EndFor
	\For{$n=1 : \SensorNum$}
	\State Sample a mini-batch of trajectories $(\mathbf{o,a},r,\mathbf{o}')$ 
	\State Update critic parameter $\omega_{n}$ using \eqref{eq:MADDPG_critic} 
	\State Update actor parameter $\theta_{n}$ using \eqref{eq:MADDPG_actor}
	\State $\omega_{n}' \gets \Softupdate \omega_{n} + (1-\Softupdate)\omega_{n}'$ 
	\State $\theta_{n}' \gets \Softupdate \theta_{n} + (1-\Softupdate)\theta_{n}'$ 
	\EndFor
	\EndFor
\end{algorithmic}
\end{algorithm}
At the beginning of each round, each sensor selects and executes the action based on its observation (Line 5).
To implement the proposed \ac{CTDE}-based algorithm, we adopt the \ac{MADDPG} algorithm \cite{LowWuTam:17}.
In this algorithm, a deterministic policy, $\mu_{\theta_{n}}$, is used to determine the action from the observation.
Here, $G_{\text{gumbel}}(\cdot)$ is Gumbel-Softmax estimator \cite{JanGuPoo:16}, which is used to support \blue{a discrete} action space. 
Next, the network status is updated and the network coverage ratio $\NetCov(t)$ is calculated (Line 7).
At the end of each round, sensors obtain the shared reward $r$; the observations of all sensors at the subsequent round $\mathbf{o}'$ are updated (Line 8); and the experience $(\mathbf{o,a},r,\mathbf{o}')$ is stored in the replay buffer (Line 9).

At the end of each episode, the mini-batch is sampled from the replay buffer, and $\omega_{n}, \theta_{n}, n \in \SensorSet$ are centrally trained at the sink node (Lines 11-13).
In order to train $\omega_{n}$, the loss for the critic network, $L(\omega_{n})$, is given by
\begin{align}\label{eq:MADDPG_critic}
L(\omega_{n})=
\mathbb{E}\left[ 
(Q_{\omega_{n}}(\mathbf{o},\mathbf{a})- r - \gamma Q_{\omega_{n}'}(\mathbf{o}',\mathbf{a}'))^{2} 
\right],
\end{align}
where $Q_{\omega_{n}}(\mathbf{o},\mathbf{a})$ denotes the action-value function, $\omega'_{n}$ is the target critic network parameters, and $\mathbf{a}'$ is the actions of all sensors at the subsequent round, respectively. 
Next, the direction of the gradient of the expected return $J(\mu_{\theta_{n}})$, $\nabla_{\theta_{n}}J(\mu_{\theta_{n}})$, is written as
\begin{align}\label{eq:MADDPG_actor}
\nabla_{\theta_{n}}J(\mu_{\theta_{n}}) =
\mathbb{E}
[\nabla_{\theta_{n}} \mu_{\theta_{n}}(a_{n}|o_{n}) 
\nabla_{a_{n}}Q_{\omega_{n}}(\mathbf{o},\mathbf{a} )|_{a_{n}=\mu_{\theta_{n}'}(o_{n})}],
\end{align}
where $\theta'_{n}$ is the target actor network parameters.
Finally, $\theta_{n}'$ and $\omega_{n}'$ are updated by adopting the soft-update with parameter $\Softupdate$ (Lines 14-15). 
Note that the proposed \ac{RL}-based \ac{SCD} algorithm also can be applied to both the single pre-charged sensor and the multiple \ac{EH} sensor cases.

\blue{Unlike tabular \ac{RL} environments, the proposed \ac{RL}-based \ac{SCD} algorithm, which is the multi-agent deep \ac{RL} algorithm, generally cannot guarantee the global optimality \cite{NadSydSim:21} nor show the performance gain theoretically.
This is because the \aratio\ is affected by sequential decisions of multiple sensors as well as the excessively many factors such as harvesting energy and channel state information during the episode.
However, this algorithm is trained to converge to a local optimum for enhancing the \aratio.
In addition, we leverage the \ac{CTDE} framework to enhance coordination among the sensors which can improve the performance. 
Moreover, we further utilize the information of the neighboring sensors by adopting the observation range.}

\subsection{\blue{Complexity Analysis}}\label{ssec:Complexity_analysis}
\blue{In this subsection, we provide the complexity analysis about the proposed \ac{RL}-based \ac{SCD} algorithm.
Here, we focus on the complexity at the sensor, which has limited computational capability and battery, because the training is processed in centralized sink node, which generally has much higher computation capability and sufficient energy.
Since the actor network is used to decide the action, the computational complexity for the sensor $n$ is affected by the size of observation, action, and the number of hidden layers and nodes \cite{CheCheWan:19, ZhaWanLiu:21}.
Hence, the complexity for execution at each round can be expressed as
}
\begin{align}\label{eq:Complexity_test}
\blue{\mathcal{O} \Big(|o_n|\HiddenNodeNum{1} + \HiddenNodeNum{1} \HiddenNodeNum{2} + \cdots + \HiddenNodeNum{\NLayerActor -1}\HiddenNodeNum{\NLayerActor} + \HiddenNodeNum{\NLayerActor} |a_n| \Big),}
\end{align}
\blue{where $|\cdot|$ is the cardinality of the set, $|a_n| = 3$ because $a_{n} \in \{\text{EC, LC, IDLE}\}$, $\HiddenNodeNum{l}$ is the number of hidden nodes of the $l$-th layer, and $\NLayerActor$ is the number of hidden layers of the actor network.
From \eqref{eq:Observation}, $|o_n|$ depends on the number of the neighboring sensors within the observation range $\ObsRange$.
Hence, the range of $|o_n|$ can be determined as}
\begin{align}\label{eq:Observation_cardinality_range}
\blue{4 \leq |o_n| \leq 3 \SensorNum + 1,}
\end{align}
\blue{which has the minimum value four when there is no neighboring sensor within the observation range, and has the maximum value when the sensor uses the information of all $\SensorNum$ sensors.}

%
\section{Simulation Results}\label{sec:Result}
In this section, we evaluate the performance of proposed probability-based and \ac{RL}-based \ac{SCD} algorithms.
%
\begin{table}[!t]
\caption{Parameter Table \label{table:parameter}} 
\begin{center}
	\rowcolors{2}
	{gray!12!}{}
	\renewcommand{\arraystretch}{1.5}
	\begin{tabular}{c|c||c|c}
		\hline 
		{\bf Description} & {\bf Value} & {\bf Description} & {\bf Value} \\
		\hline\hline
		$\SlotDuration$ & $10$ ms & $\Bandwidth$ & $10$ MHz \\
		\hline
		$\ProcSlotE$ & $1$ & $\ProcSlotL$ & $2$ \\
		\hline
		$\sigma^2$ & $-100$ dBm & $\BSdensity$ & $10^{-4}$ nodes/$\text{m}^{2}$ \\
		\hline
		$\Ptx{n}$ & $15$ dBm & $\PathExp$ & $4$\\
		\hline
		$\ErrTh$ & $0.6$ & $\RoundLen$ & $8$ \\
		\hline
		$\InDataSize$ & $6$ Kbit & $\OutDataSize$ & $500$ bit \\
		\hline
		$\beta_{1}$ & $0.0045$ & $\beta_{2}$ & $1.35$ \\
		\hline
		$T_{\text{ep}}$ & $20$ & $\ObsRange$ & $100$ m \\
		\hline
		$\gamma$ & $0.95$ & $T_{\text{train}}$ & $20,000$ \\
		\hline
		$\Softupdate$ & $0.005$ & $\RewardPenalty$ & $1$ \\
		\hline 
	\end{tabular}
\end{center}
\end{table}%
%
We set $\TotRound = 20$, $\Txcount = 3$, $\E{\SE} = 10$ (mJ), $\Et = 13.55$ (mJ) \cite{GonCheMa:18}, \blue{$p_\text{t} = 1$} and $\TgtCov = 0.9$.
For the single pre-charged sensor case, we consider the network $\CoordinateSingle$ as a circle with the radius of 50 (m), and the sensor is placed at a center of the network. 
In this case, $\Ep = 12$ (mJ) and the battery constraint during $\TotRound$ rounds is $\PreBatteryTh = 400$ (mJ).
For multiple \ac{EH} sensor case, we consider the network $\Coordinate$ as 250 m $\times$ 250 (m) square area where the sensors are irregularly placed in the network.
In this case, $\SensorNum=10$, $\Ep = 20$ (mJ), $\Bmax =50$ (mJ) and harvested energy during one time slot is $\HarvEnergy{n}{i} \sim \mathcal{U}[1.5, 4.5]$ (mJ), $\forall n,i$ with the average harvesting energy $\E{\Harv} = 3$ (mJ) \cite{LuoPuWan:19}. 
In this case, we compare the performance of the following algorithms.

\begin{itemize}
\item \textbf{Probability-based SCD (Probability-SCD)}: This is the algorithm to make the \ac{SCD} with certain probabilities \cite{KumLaiBal:04,LiaPenHua:20}. 
For this case, the optimal sensing and computing probabilities that maximize the \aratio, provided in Section \ref{sec:Probability_SCD}, are used.
\item \textbf{RL-based SCD (RL-SCD)}:
This is the proposed algorithm that makes the \ac{SCD} at sensors based on the policy trained by \textbf{Algorithm} in Section \ref{sec:MARL_SCD}.
\item \textbf{\ac{RL}-based sensing decision with \ac{EC} (RL-SD (EC))}: 
Sensors make sensing decisions based on policy trained by the \ac{RL} algorithm, and the generated data is computed at the \ac{EC} server. 
\item \textbf{\ac{RL}-based sensing decision with \ac{LC} (RL-SD (LC))}: Sensors make sensing decisions based on policy trained by the \ac{RL} algorithm, and the generated data is locally computed. 
\item \textbf{RL-based SCD with confident information coverage (RL-SCD (CIC)) \cite{WanDenLiu:13}}: Sensors make \ac{SCD} based on the policy trained by the \ac{RL} algorithm, but the temporal correlation of information is not considered, as in \cite{WanDenLiu:13}. 
The sensing coverage becomes a circle with radius $r_{\text{CIC}} = \SensingRad(\RoundLen\SlotDuration, \ErrTh)$ from the instant of the successful update until the next round, where $\SensingRad(\RoundLen\SlotDuration, \ErrTh)$ is the sensing radius at the end of the current round, defined in \eqref{eq:Sensing_radius}.
\end{itemize}

To train the sensors, we use an actor-critic neural network, \blue{where both networks have two fully connected hidden layers with 64 neurons for each layer, i.e., $\NLayerActor=\NLayerCritic=2$, $\HiddenNodeNum{1} = \HiddenNodeNum{2} = 64$.\footnote{\blue{Due to the simple network structure with a small number of nodes and layers, the sensor can make decisions with low complexity through the actor network.}}
Here, $\NLayerCritic$ is the number of hidden layers of critic network.
In addition, we use Adam optimizer with mini-batch size as 512.}
Other parameters used in the simulation are shown in Table \ref{table:parameter}.

\begin{figure}[!t]
\centering
\psfrag{Aratio}[Bc][Tc][0.7]{$\TgtCov$-coverage probability, $\ECovProb$}
\psfrag{Eta}[Tc][Bc][0.7]{Target coverage ratio, $\TgtCov$}
\psfrag{MARL-based-DELTAONEDDDDDDDDDDDDD}[tl][tl][0.65]{RL-SCD, $\Txcount=1$}
\psfrag{MARL-based-DELTATHREE}[tl][tl][0.65]{RL-SCD, $\Txcount=3$}
\psfrag{Probability-DELTAONE}[tl][tl][0.65]{Probability-SCD (Analysis), $\Txcount=1$}
\psfrag{Probability-DELTATHREE}[tl][tl][0.65]{Probability-SCD (Analysis), $\Txcount=3$}
\psfrag{Simulation1}[tl][tl][0.65]{Probability-SCD (Simulation), $\Txcount=1$}
\psfrag{Simulation2}[tl][tl][0.65]{Probability-SCD (Simulation), $\Txcount=3$}
\psfrag{Optimal-prob}[Bc][Tc][0.8]{$\Ps^{*}, \Pe^{*}$}
\psfrag{Aratio-analysis}[tl][tl][0.65]{$\ECovProb$ (Analysis)}
\psfrag{Aratio-simulationDDD}[tl][tl][0.65]{$\ECovProb$ (Simulation)}
\psfrag{Opt-ps}[tl][tl][0.65]{$\Ps^{*}$}
\psfrag{Opt-pe}[tl][tl][0.65]{$\Pe^{*}$}
\psfrag{Distance}[Tc][Bc][0.7]{Distance between the sensor - sink node, $d_{\textbf{x}_{1},\text{o}}$}
\subfigure[\aratio\ as a function of $\TgtCov$ for \blue{different} $\Txcount$ and $\PreBatteryTh$ when $d_{\textbf{x}_{1},\text{o}}$ = 100 (m).]{\includegraphics[width=0.9\columnwidth]{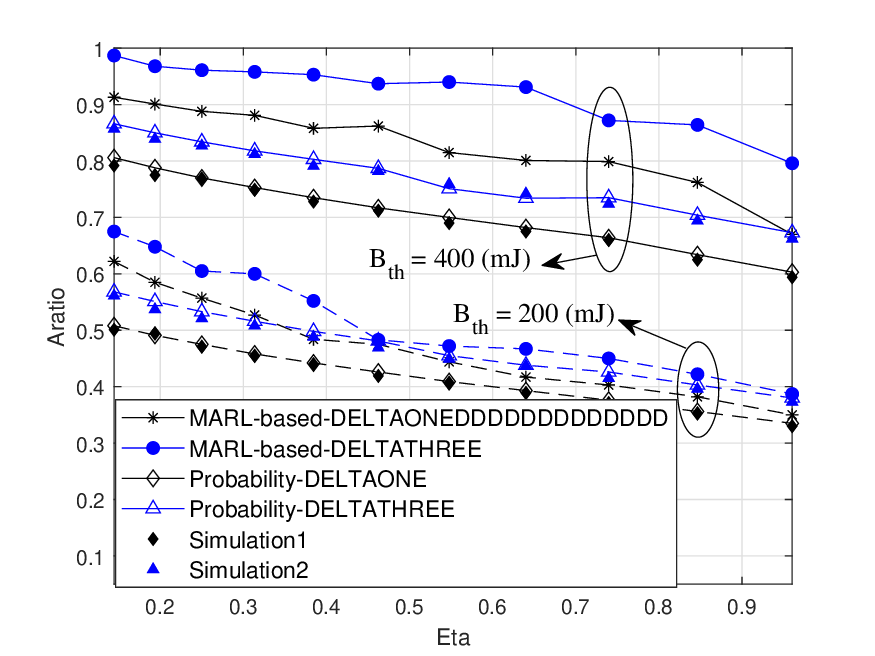}
	\label{fig:Single_precharged_vth}
}
%
\subfigure[\aratio\ and the optimal sensing and \ac{EC} probabilities $\Ps^{*}, \Pe^{*}$ of Probability-SCD as a function of $d_{\textbf{x}_{1},\text{o}}$.]{\includegraphics[width=0.9\columnwidth]{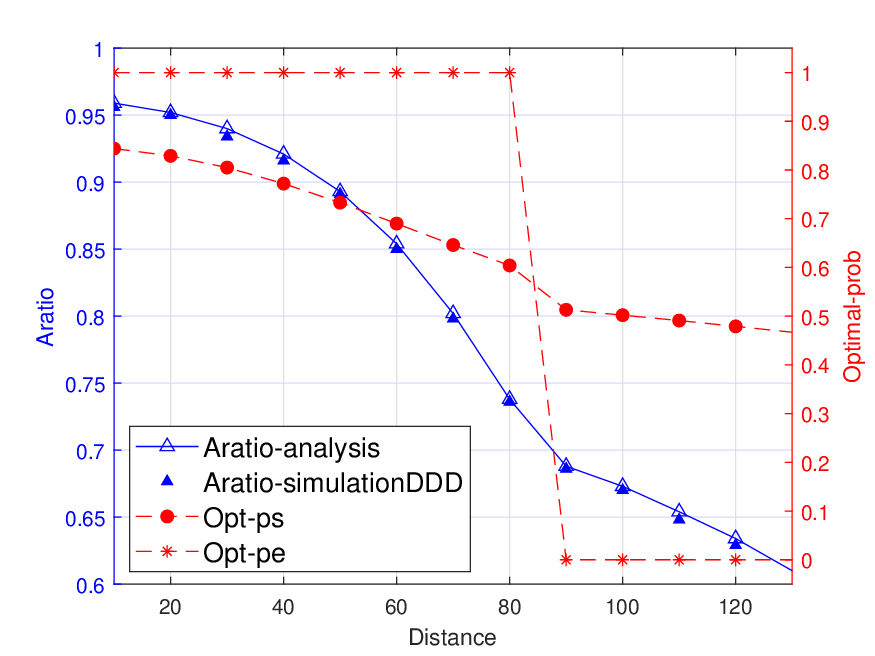} 
	\label{fig:Single_precharged_dist}}
\caption{\aratio\ for the single pre-charged sensor case.}
\label{fig:Result_single_precharged}
\end{figure}
First, we show the \aratio\ of the single pre-charged sensor case in Fig. \ref{fig:Result_single_precharged}.
Figure \ref{fig:Single_precharged_vth} shows the \aratio\ as a function of the target coverage ratio, $\TgtCov$, for different maximum number of transmission attempts $\Txcount$ and the battery constraint $\PreBatteryTh$ when $d_{\textbf{x}_{1},\text{o}}$ = 100 (m).
From Fig. \ref{fig:Single_precharged_vth}, we first see that the simulation result of Probability-SCD matches well with our analysis, described in Section \ref{ssec:prob_SCD_single} and Remark 1.
In this figure, the \aratio\ decreases with $\TgtCov$ due to the decrease of the target \ac{AoI} $\vth$, as shown in \eqref{eq:Aratio_AVP}.
In addition, the \aratio\ is higher when $\PreBatteryTh$ = 400 (mJ) compared to that when $\PreBatteryTh$ = 200 (mJ) because the sensor can perform sensing more frequently with higher $\PreBatteryTh$.
Moreover, the \aratio\ is higher with $\Txcount = 3$ compared to that with $\Txcount = 1$ because the data can reliably reach the sink node by adopting more retransmission opportunities, which is helpful in maintaining the data freshness.
Note that RL-SCD outperforms the Probability-SCD as it utilizes the current \ac{AoI} and battery level for the sensing and computing decision.

\blue{Figure \ref{fig:Single_precharged_dist} shows the \aratio\ with the optimal sensing and EC probabilities, $\Ps^{*}$ and $\Pe^{*}$, of Probability-SCD.
Specifically, we show the effects of the distance between the sensor and the sink node, $d_{\textbf{x}_{1},\text{o}}$, on the performance of Probability-SCD.}
When $d_{\textbf{x}_{1},\text{o}}$ is short, the outage probability is low, as shown in \blue{\eqref{eq:Outage_prob_result}}, and the average number of transmissions is reduced, which decreases $\bar{E}_\text{e}$ in \eqref{eq:Energy_edge} and $\bar{E}_\loc$ in \eqref{eq:Energy_local}.
In addition, $\bar{E}_\text{e} < \bar{E}_\loc$ because the sensor does not consume the computing energy when the sensed data is computed at the \ac{EC} server.
Therefore, from Fig. \ref{fig:Single_precharged_dist}, $\Pe^{*}$ is equal to one when $d_{\textbf{x}_{1},\text{o}} \leq 80$ (m).
However, when $d_{\textbf{x}_{1},\text{o}} > 80$ (m), from \blue{\eqref{eq:Outage_prob_result}}, \ac{EC} is more likely to have larger increase in outage probability and the energy consumption due to the larger data size to transmit, compared to \ac{LC}, so $\Pe^{*}$ becomes zero.
Here, $\Pe^{*}$ becomes one or zero in Fig. \ref{fig:Single_precharged_dist} because only the communication and computation performance affect the \aratio\ and the waiting time at the \ac{EC} server is negligible in the single pre-charged sensor case, as explained in Section \ref{ssec:prob_SCD_single}.
Additionally, since $\bar{E}_\text{e}$ and $\bar{E}_\loc$ increase with $d_{\textbf{x}_{1},\text{o}}$, from \eqref{eq:Ps_opt_for_pe}, $\Ps^{*}$ also decreases with $d_{\textbf{x}_{1},\text{o}}$.
Accordingly, we can also see that the \aratio\ decreases with $d_{\textbf{x}_{1},\text{o}}$ because the sensor cannot perform sensing frequently for low $\Ps^{*}$.

\begin{figure}[t!]
\centering
\subfigure[Training curve of proposed RL-SCD algorithm for different penalty factor $\RewardPenalty$.]{\includegraphics[width=0.9\columnwidth]{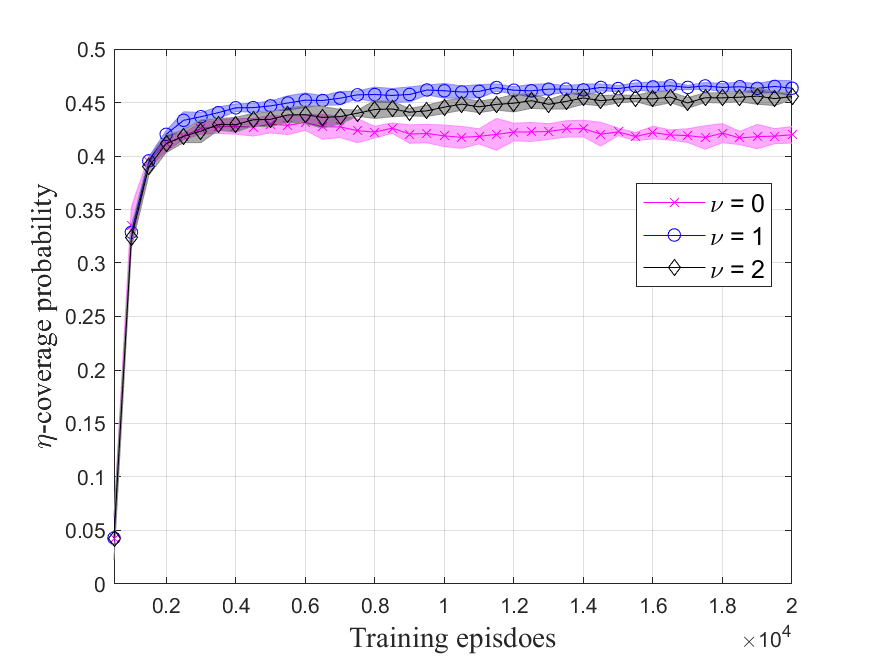}
	\label{fig:Result_penalty_factor}
}
\subfigure[Training curve of the proposed RL-SCD algorithm and baseline algorithms.]{\includegraphics[width=0.9\columnwidth]{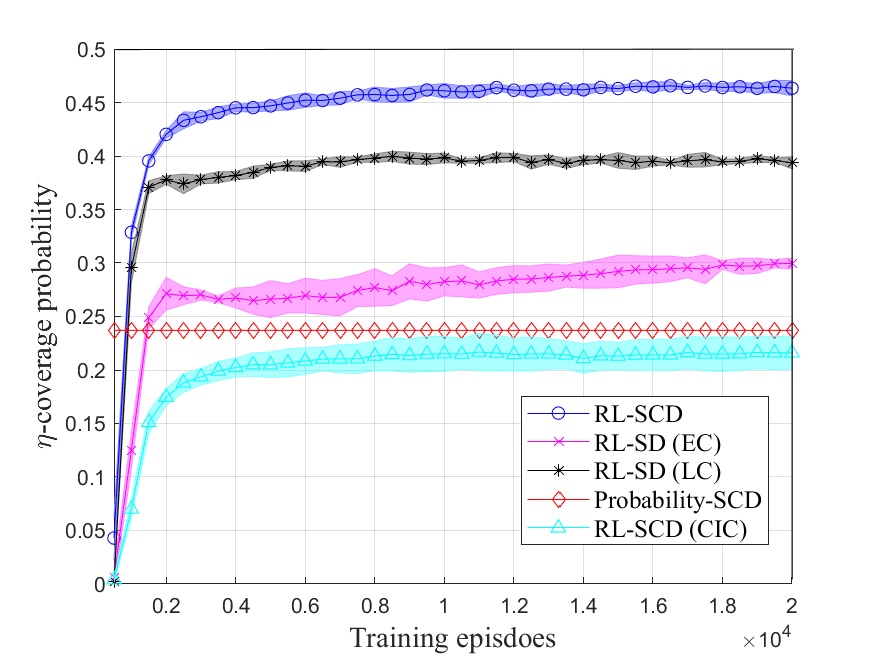}
	\label{fig:Training_curve_algorithms}
}
\caption{
	Training curve for the multiple \ac{EH} sensor case.
}
\label{fig:Training_curve}
\end{figure}

\begin{figure}[!t]
\centering
\psfrag{Aratio}[Bc][Tc][0.8]{$\TgtCov$-coverage probability, $\ECovProb$}
\psfrag{Sensornum}[tc][tc][0.8]{The number of sensors, $\SensorNum$}
\psfrag{MARL-based-DDDDDD}[tl][tl][0.7]{RL-SCD}
\psfrag{MARL-EC}[tl][tl][0.7]{RL-SD (EC)}
\psfrag{MARL-LC}[tl][tl][0.7]{RL-SD (LC)}
\psfrag{Prob}[tl][tl][0.7]{Probability-SCD}
\psfrag{CIC}[tl][tl][0.7]{RL-SCD (CIC)}
\psfrag{Eta85}[tl][tl][0.65]{$\TgtCov=0.85$}
\psfrag{Eta90}[tl][tl][0.65]{$\TgtCov=0.9$}
\psfrag{EightSensors}[tc][tc][0.8]{$\SensorNum=8$}
\psfrag{FourteenSensors}[tc][tc][0.8]{$\SensorNum=14$}
\psfrag{1004}[tc][tc][0.65]{$\mathbb{E}_{t}[\NetCov(t)]$}
\psfrag{2004}[tc][tc][0.65]{Sensing ratio}
\psfrag{3004}[tc][tc][0.65]{EC ratio}
\psfrag{4004}[tc][tc][0.65]{$\mathbb{E}_{n,t}[\SinkAoI{n}{t}]$(ms)}
\subfigure[\aratio\ of proposed and baseline algorithms according to \blue{$\SensorNum$ for different} $\TgtCov$.]{\includegraphics[width=0.9\columnwidth]{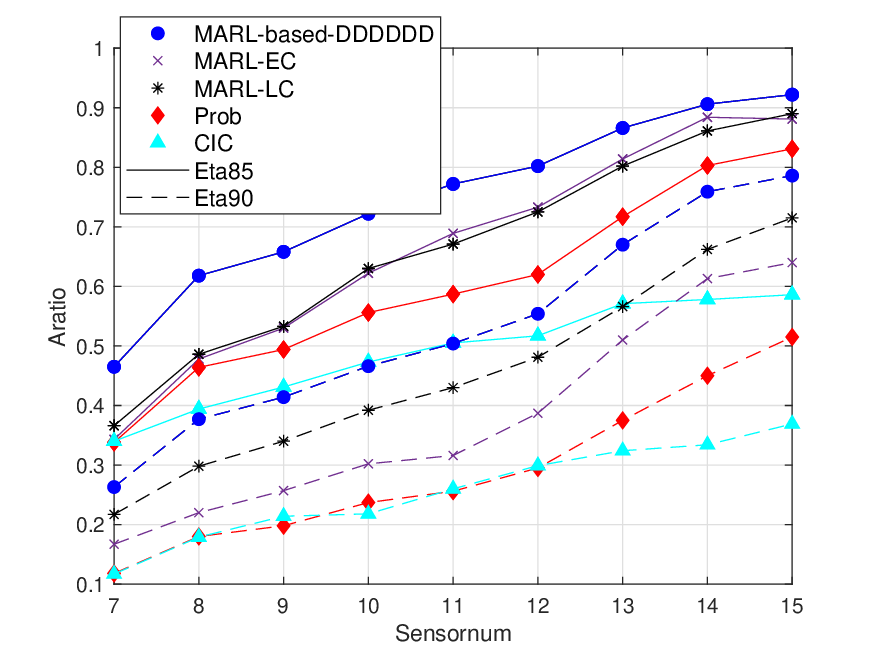}
	\label{fig:sensor_num_aprob}
}
\subfigure[{Average network coverage ratio $\mathbb{E}[\NetCov(t)]$, sensing ratio, and \ac{EC} ratio of RL-SCD and Probability-SCD for $\SensorNum=8$ and $\SensorNum=14$ when $\TgtCov=0.9$.}]{\includegraphics[width=0.8\columnwidth]{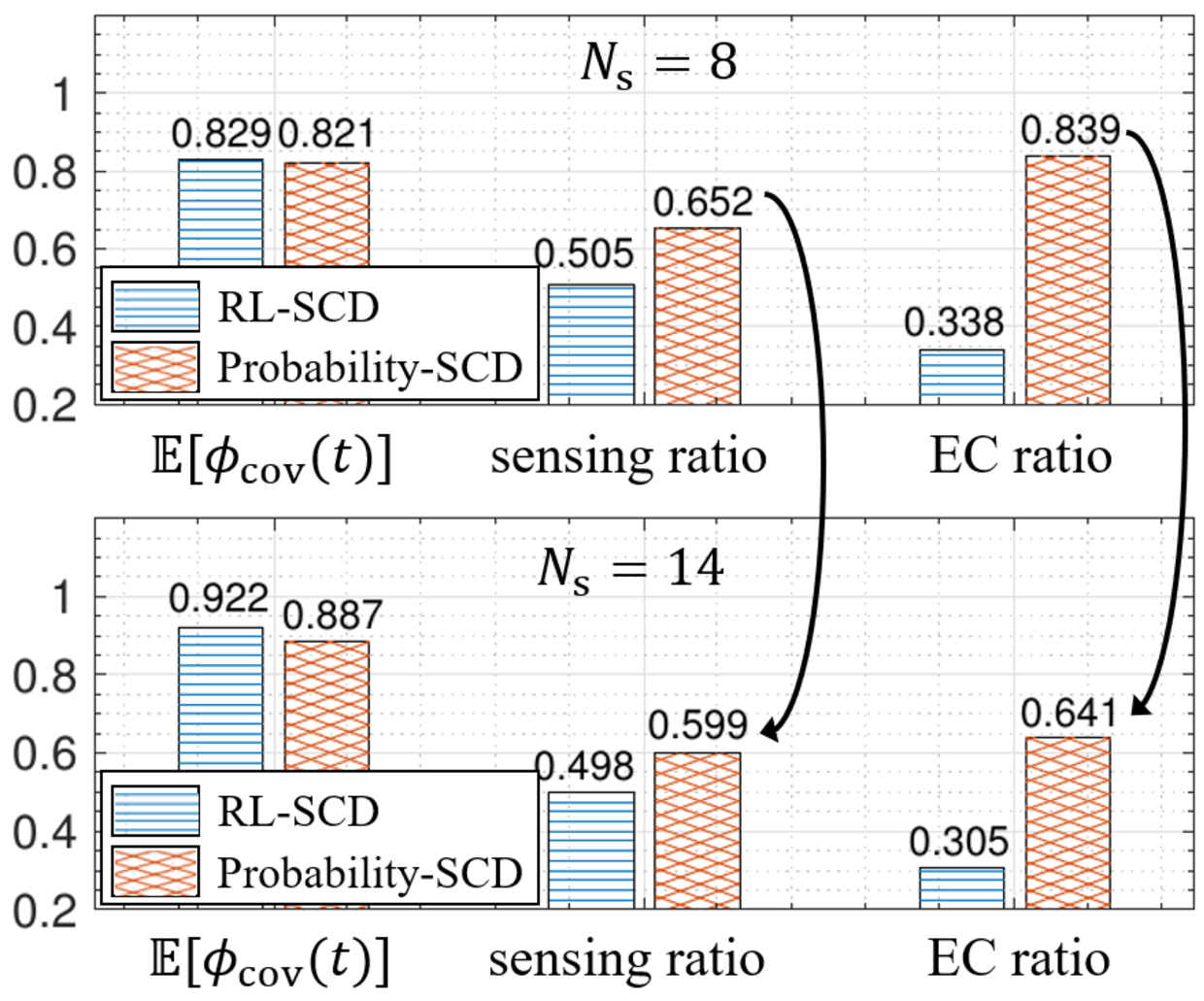}
	\label{fig:sensor_num_ratio}
}
\caption{	
	Effect of the number of sensors, $\SensorNum$, on the network performance. 
}
\label{fig:multiple_sensor_num}
\end{figure}

\begin{figure}[!t]
\centering
\psfrag{Aratio}[Bc][Tc][0.8]{$\TgtCov$-coverage probability, $\ECovProb$}
\psfrag{ProcEnergy}[tc][tc][0.8]{Computing energy, $\Ep$ (mJ) }
\psfrag{MARL-based-DDDDDD}[tl][tl][0.7]{RL-SCD}
\psfrag{MARL-EC}[tl][tl][0.7]{RL-SD (EC)}
\psfrag{MARL-LC}[tl][tl][0.7]{RL-SD (LC)}
\psfrag{Prob}[tl][tl][0.7]{Probability-based}
\psfrag{CIC}[tl][tl][0.7]{RL-SCD (CIC)}
\psfrag{LowDensity}[tl][tl][0.65]{$p_{\text{t}}=0.5$}
\psfrag{HighDensity}[tl][tl][0.65]{$p_{\text{t}}=1$}$  $
\psfrag{ProcEnergy12}[tc][tc][0.8]{$\Ep = 12$ (mJ)}
\psfrag{ProcEnergy28}[tc][tc][0.8]{$\Ep = 28$ (mJ)}
\psfrag{1004}[tc][tc][0.65]{$\mathbb{E}_{t}[\NetCov(t)]$}
\psfrag{2004}[tc][tc][0.65]{Sensing ratio}
\psfrag{3004}[tc][tc][0.65]{EC ratio}
\psfrag{4004}[tc][tc][0.65]{$\mathbb{E}_{n,t}[\SinkAoI{n}{t}]$(ms)}
\subfigure[\aratio\ of proposed and baseline algorithms according to \blue{$\Ep$ for different resource reuse probability $p_{\text{t}}$.}]{\includegraphics[width=0.9\columnwidth]{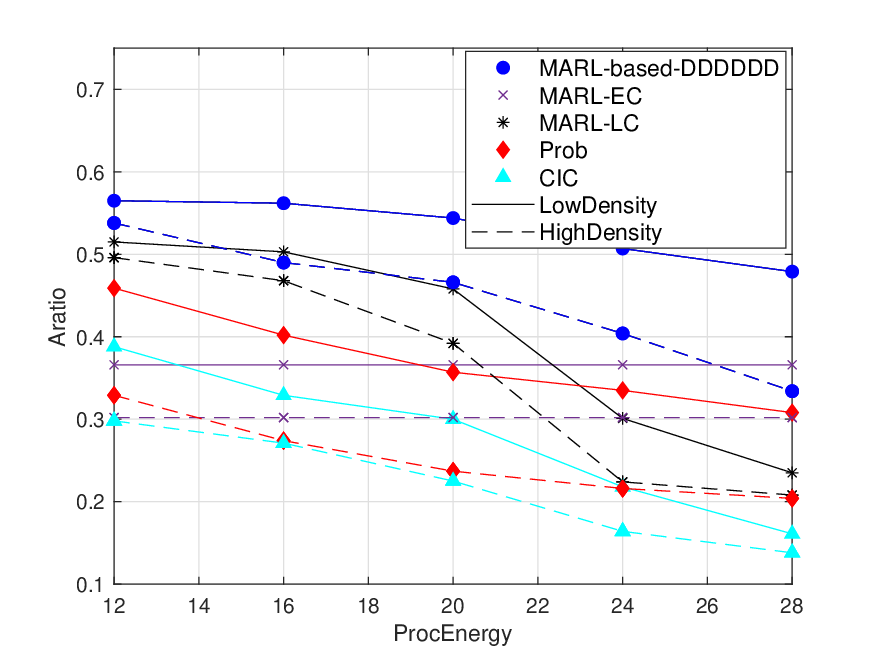}
	\label{fig:proc_energy_aprob}
}
\subfigure[{Average network coverage ratio $\mathbb{E}[\NetCov(t)]$, sensing ratio, and \ac{EC} ratio of RL-SCD and Probability-SCD for $\Ep = 12$ mJ and $\Ep = 28$ mJ \blue{when $p_{\text{t}}=0.5$.}}]{\includegraphics[width=0.8\columnwidth]{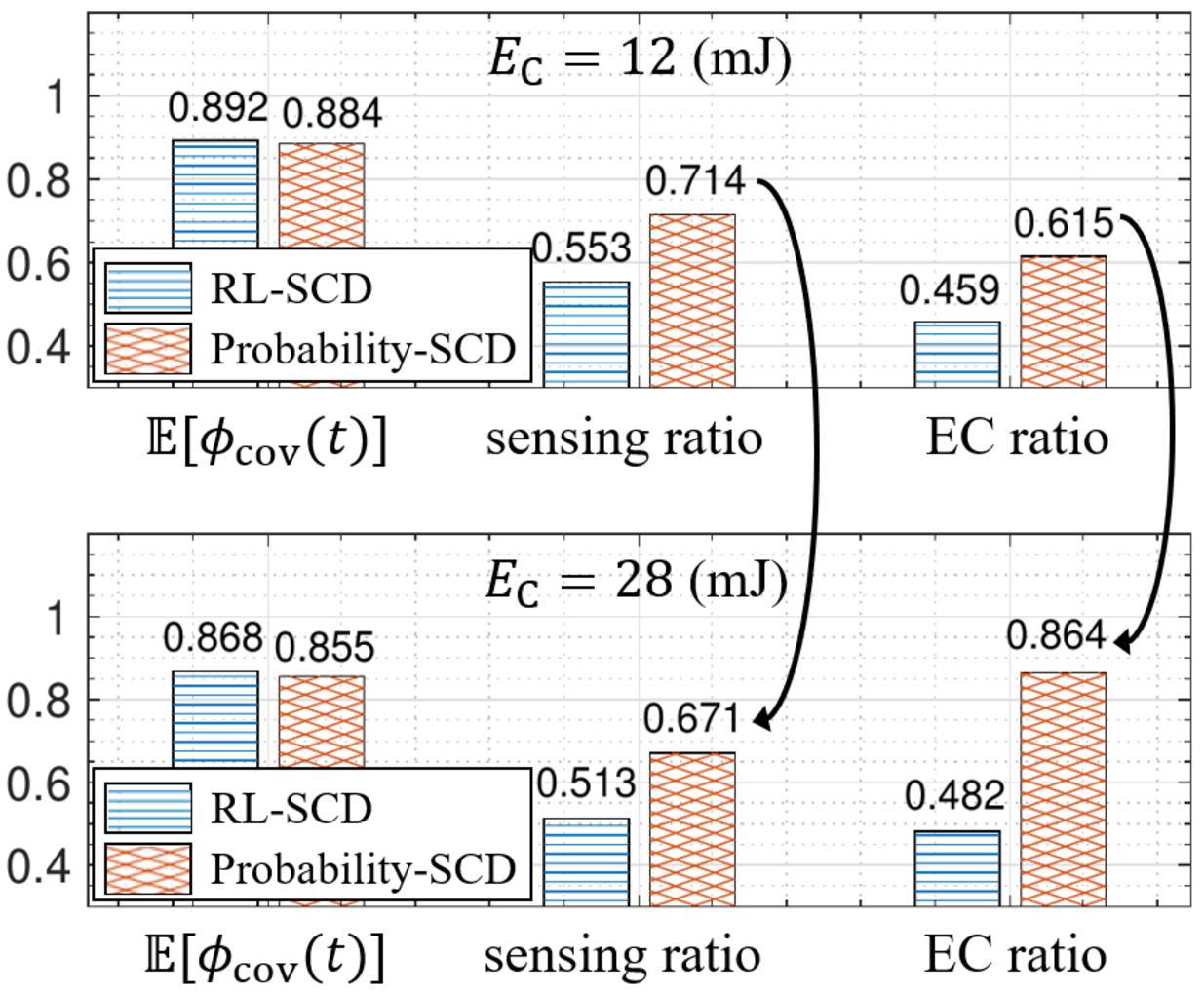}
	\label{fig:proc_energy_ratio}
}	
\caption{	
	Effect of the computing energy, $\Ep$, on the network performance. 
}
\label{fig:proc_energy}
\end{figure}

Next, we show the simulation results of the multiple \ac{EH} sensor case in Figs. \ref{fig:Training_curve} - \ref{fig:proc_energy}.
Figure \ref{fig:Training_curve} shows training curves, averaged over different seeds. 
In Fig. \ref{fig:Result_penalty_factor}, the proposed RL-SCD algorithm can achieve the higher \aratio\ \blue{with the penalty factor} $\RewardPenalty=1$, compared to those with $\RewardPenalty=0$ and $\RewardPenalty=2$, so we adopt $\RewardPenalty=1$ for implementing the RL-based algorithms.
Figure \ref{fig:Training_curve_algorithms} presents the training curves of the proposed RL-SCD and baseline algorithms.
From this figure, we observe that the \aratio\ increases and converges as the training proceeds for RL-based algorithms.
Note that the Probability-SCD is not a learning-based algorithm, there is no training curve.
However, we present the \aratio\ for this case as a line for the performance comparison purpose.
We can observe that the proposed RL-SCD achieves higher performance than other algorithms. 
From Fig. \ref{fig:Training_curve_algorithms}, we can see that the RL-SCD (CIC) has lower performance even than Probability-SCD, which shows the importance of exploiting the temporal correlation of information for the sensing and computing decision.

Figure \ref{fig:multiple_sensor_num} shows the effect of the number of sensors, $\SensorNum$, on the network performance.
Firstly, Fig. \ref{fig:sensor_num_aprob} shows the \aratio\ as a function of $\SensorNum$ for \blue{different} $\TgtCov$.
From this figure, we find that the \aratio\ increases with $\SensorNum$ because more sensors can cover larger area, as shown in \eqref{eq:Coverage_point}.
We can also observe that the proposed algorithm can achieve \blue{the higher} \aratio\ compared to the baseline algorithms even for high $\TgtCov$.

In Fig. \ref{fig:sensor_num_ratio}, we present the average network coverage ratio $\left(\mathbb{E}[\NetCov(t)]\right)$, the ratios of performing (deciding) sensings or \ac{EC} in each round, when $\SensorNum=8$ and $\SensorNum=14$.
For Probability-SCD, we use the optimal $\Ps^{*}$ and $\Pe^{*}$, obtained in Section \ref{sec:Probability_SCD}, which are the same as the obtained sensing ratio and the \ac{EC} ratio, respectively, in this case.
\blue{Firstly, we observe that when the number of sensors is changed from $\SensorNum=8$ to $\SensorNum=14$, both the sensing ratio and the \ac{EC} ratio become lower, to reduce the excessive energy consumption and the computation load at the \ac{EC} server.}
From this figure, we can see that the \ac{EC} ratio of RL-SCD is smaller than that of Probability-SCD. 
This is because, in RL-SCD, the \ac{AoI} and the waiting time at the \ac{EC} server are used as an observation to avoid the excessive waiting at the \ac{EC} server.
From Fig. \ref{fig:sensor_num_ratio}, we can also see that the sensing ratio of RL-SCD is lower than that of Probability-SCD, which shows RL-SCD can achieve higher coverage ratio even with fewer sensing.

Figure \ref{fig:proc_energy} shows the effect of the computing energy, $\Ep$, on the network performance when $\SensorNum = 10$.
Figure \ref{fig:proc_energy_aprob} shows the \aratio\ of \blue{the proposed} and baseline algorithms as a function of $\Ep$ for \blue{different resource reuse probability $p_{\text{t}}$}.
From Fig. \ref{fig:proc_energy_aprob}, we can see that the \aratio\ decreases as $\Ep$ increases, except for RL-SD (EC), which is not affected by $\Ep$ as it does not perform LC.
\blue{Moreover, the \aratio\ becomes lower with $p_{\text{t}}=1$ compared to that with $p_{\text{t}}=0.5$ because higher $p_{\text{t}}$ means larger interference from other sensors, which increases the outage probability.}
However, \ac{RL}-SCD can still achieve \blue{the higher} \aratio\ than other algorithms for all ranges of $\Ep$ in Fig. \ref{fig:proc_energy}.

Figure \ref{fig:proc_energy_ratio} shows the average network coverage ratio $\mathbb{E}[\NetCov(t)]$, the sensing ratio, and the \ac{EC} ratio when $\Ep = 12$ (mJ) and $\Ep = 28$ (mJ).
From Fig. \ref{fig:proc_energy_ratio}, we can see that when $\Ep$ is changed from 12 (mJ) to 28 (mJ), the sensing ratio becomes lower and the \ac{EC} ratio becomes higher\blue{, to prevent the excessive energy consumption for sensing and computing.}


\section{Conclusion}\label{sec:Conclusion}
This paper proposes the probability-based \ac{SCD} and the \ac{RL}-based \ac{SCD} algorithms for data freshness in the \ac{EC}-enabled wireless sensor networks.
Firstly, we propose the novel sensing coverage which is jointly affected by the spatial and the temporal correlation of information.
Then, we design the probability-based \ac{SCD} algorithm and derive the \aratio\ in a closed form in the \blue{single pre-charged sensor} case. Next, we obtain the optimal sensing and \ac{EC} probabilities that maximize the \aratio.
We also design the \ac{RL}-based \ac{SCD} algorithm by training the policy of the sensors to take the real-time decision based on its observation in both the single pre-charged sensor and the multiple \ac{EH} sensor cases. 
Through the simulation results, we first show the \ac{RL}-based \ac{SCD} algorithm can achieve \blue{the higher} \aratio\ than other baseline algorithms.
We also provide some insights on the \ac{SCD}, i.e., in the single pre-charged sensor case, as the distance between the sensor and the sink node increases, choosing the \ac{LC} rather than \ac{EC}, is helpful to enhance the \aratio. 
In addition, in the multiple \ac{EH} sensor case, we find that both the sensing ratio and \ac{EC} ratio decrease with the number of sensors, 
while \blue{the \ac{EC} ratio increases} with the computing energy.

\begin{appendix}
\subsection{Proof of Lemma 1}\label{app:Lemma_1}
The expected inter update time $\Uk$ is represented as 
\begin{align}\label{eq:Uk_def}
	\Uk = \mathbb{E} \left[ \RoundLen - Z_{k-1} + X_k \right]
	\mathrel{\overset{\makebox[0pt]{{\tiny(a)}}}{=}} \RoundLen - \Zk + \Xk,
\end{align}
where $Z_k$ is the time interval from the sensing instance to the successful update, and $X_k$ is the time interval from the first sensing decision time after the $(k-1)$-th successful update to the $k$-th successful update. 
Here, (a) holds because\textit{ 1) $(k-1)$-th update and $k$-th update are independent} and \textit{2) $Z_{k-1}$ and $Z_{k}$ are identically distributed.}

Since $\Zk$ and $\Xk$ are affected by what computing model is used for the $k$-th successful update, i.e., $\UpdateK \in \{\text{e}, \loc\}$,
both terms are expressed as the sum of conditional expectations $\mathbb{E}[Z_k  | \UpdateK]$ and $\mathbb{E}[X_k  | \UpdateK]$ as follows 
\begin{align} \label{eq:Zk_def}
	\Zk  =   \Peprime{\text{e}} \mathbb{E}[Z_k|\text{e}] 
	+  \Peprime{\loc} \mathbb{E}[Z_k|\loc], 
\end{align}
\begin{align}\label{eq:Xk_def}
	\Xk  =   \Peprime{\text{e}} \mathbb{E}[X_k|\text{e}] 
	+ \Peprime{\loc} \mathbb{E}[X_k|\loc].
\end{align}
%

We first obtain $\mathbb{E}[Z_k|\UpdateK]$.
For given $\UpdateK$, $Z_k$ is determined by the number of retransmissions attempted by the sensor until the transmission succeeds, denoted as $c$, so its distribution is modeled as the geometric distribution.
$\mathbb{P}[Z_k|\UpdateK]$ is given by
\begin{align}\label{eq:Zk_pmf}
	\mathbb{P}[Z_k = 1 + c + \tk|\UpdateK] 
	= \frac{\OutageProb{k}^{c-1} \TxProbxTime{k}{1}}
	{\TxProbPf{k}}.
\end{align}
From \eqref{eq:Zk_pmf}, $\mathbb{E}[Z_k|\UpdateK]$ is then obtained as
\begin{align}\label{eq:Zk_cndt}
	\begin{aligned}
		\mathbb{E}[Z_k|\UpdateK] 
		= \sum_{c=1}^{\Txcount}{(1+c+\tk)} 
		\frac{\OutageProb{k}^{c-1} \TxProbxTime{k}{1}}
		{\TxProbPf{k}} = 1+\tk + \FuncOne{k}{\Txcount},
	\end{aligned}
\end{align}
where $\FuncOne{k}{\Txcount}$ is defined at \eqref{eq:FuncOne}.
From this, the expectation of the time interval from the sensing instance to the successful update $\Zk$ can be obtained.

Similarly, we obtain $\mathbb{E}[X_k|\UpdateK]$. 
For given $\UpdateK$, $X_k$ is affected by the number of rounds until the $k$-th update occurs, denoted as $Y_k$.
When the $k$-th update occurs from $(k-1)$th update to $Y_k$ rounds after, the \ac{PMF} $\mathbb{P}[Y_k = y|\UpdateK]$ is given by
\begin{align}\label{eq:Yk_pmf}
	\begin{aligned}
		\mathbb{P}\left[Y_k = y |\UpdateK\right]  
		= \Px (1 - \Px ) ^{y - 1}.
	\end{aligned}
\end{align}
The conditional expectation $\mathbb{E}[X_k|\UpdateK]$ is then obtained as 
\begin{align}\label{eq:Xk_cndt}
	\begin{aligned}
		\mathbb{E}[X_k|\UpdateK] =& \sum_{y=1}^\infty \mathbb{P}[Y_k = y|m_k]\{ (y-1)\RoundLen + \mathbb{E}[Z_k|\UpdateK]\}  \\
		= & \frac{\RoundLen}{\Px}- \RoundLen +  1+\tk + \FuncOne{k}{\Txcount}. \\
	\end{aligned}
\end{align}
By substituting \eqref{eq:cndt_edge_prob}, \eqref{eq:Zk_cndt} and \eqref{eq:Xk_cndt} into \eqref{eq:Zk_def} and \eqref{eq:Xk_def}, the expected inter-update time $\Uk$ can be obtained as \eqref{eq:Uk_result}.

\subsection{Proof of Lemma 2}\label{app:Lemma_2}
To show the proof of the Lemma 2, we define the probability that the target \ac{AoI} $\vth$ exists in the given ranges below, represented as 
\begin{align}\label{eq:vth_range}
	\POne &= \mathbb{P}[\AVPCaseOne | \UpdateKMO], \\
	\PTwo &= \mathbb{P}[\AVPCaseTwo | (\UpdateKMO, \UpdateK)], \\
	\PThree &= \mathbb{P}[\AVPCaseThree | (\UpdateKMO, \UpdateK)]. 
\end{align}

From \eqref{eq:gk_def}, $g_k$ is jointly affected by $\UpdateKMO$ and $\UpdateK$.
Hence, as shown in \eqref{eq:E_gk_def}, we analyze the $\gk$ by splitting it into $\UpdateKMO$ and $\UpdateK$, i.e.,  $G(\UpdateKMO, \UpdateK) = \mathbb{E}[g_k | (\UpdateKMO,\UpdateK)]$, which is expressed as
\begin{align}\label{eq:gk_cndt_def}
	\begin{aligned}
		G(\UpdateKMO, \UpdateK) \\
		= \POne \mathbb{E}&[\RoundLen - Z_{k-1} + X_k | \AVPCaseOne, (\UpdateKMO, \UpdateK)] \\
		+ \PTwo \mathbb{E}&\left[\RoundLen + X_k - (\vfloor+1) | \right. \\
		&\left.	\AVPCaseTwo, (\UpdateKMO, \UpdateK)\right] .
	\end{aligned}
\end{align}
Since \eqref{eq:gk_cndt_def} depends on the range of $\vth$,
$G(\UpdateKMO, \UpdateK)$ is obtained by following cases. 

For $\vth < 2+\tkmo$, the \ac{AoI} at the sink node $\SinkAoISgl{t}$ always violates $\vth$ 
because the minimum time for the update with the sensed data is $2+\tkmo$ when the transmission succeeds at once, i.e., $c=1$.
Hence, $\vth$ is always lower than $Z_{k-1}$, and we have
\begin{align}\label{eq:Case1_P3}
	\begin{aligned}
		\POne = 1, \PTwo = 0, \PThree = 0.
	\end{aligned}
\end{align}
By substituting \eqref{eq:Case1_P3} into \eqref{eq:gk_cndt_def},  $G(\UpdateKMO, \UpdateK)$ is obtained as
\begin{align}\label{eq:Case1_result}
	\begin{aligned}
		G(\UpdateKMO, \UpdateK)& = \mathbb{E}[\RoundLen - Z_{k-1} + X_k|(\UpdateKMO, \UpdateK)]\\
		& = \RoundLen - \mathbb{E}[Z_{k-1}| \UpdateKMO] + \mathbb{E}[X_k| \UpdateK],
		%
	\end{aligned}
\end{align}	
where the result is given in \eqref{eq:gk_cndt_result_onecol}, case $\vth < v_1$.

For $2+\tkmo \leq \vth < 1 + \Txcount + \tkmo$, the maximum number of retransmissions is limited by $\vfloor -1 - \tkmo$ to satisfy $\vth \geq Z_{k-1}$.
Hence, we have 
\begin{align}\label{eq:Case2_P}
	\begin{aligned}
		\PTwo &= \sum_{c=1}^{\vfloor -1 - \tkmo}
		\frac{\OutageProb{k-1}^{c-1} \TxProbxTime{k-1}{1}}
		{\TxProbPf{k-1}} = \frac{\TxProbxTime{k-1}{\vfloor -1 - \tkmo}}
		{\TxProbPf{k-1}},\\
		\POne &= 1 - \PTwo, \qquad \PThree = 0.
	\end{aligned}
\end{align}
Since $\POne + \PTwo = 1$, \eqref{eq:gk_cndt_def} can be rewritten as
\begin{align}\label{eq:Case2_rewritten}
	\begin{aligned}
		G(\UpdateKMO, \UpdateK)  
		&= \RoundLen + \mathbb{E}[X_k|\UpdateK] - \PTwo(\vfloor + 1) \\
		&- \POne\mathbb{E}[Z_{k-1}|\AVPCaseOne,\UpdateKMO].
	\end{aligned}
\end{align}
Here, $\mathbb{E}[Z_{k-1}|\AVPCaseOne,\UpdateKMO]$ is represented as
\begin{align}\label{eq:Case2_ZvleqZ}
	\begin{aligned}
		&\mathbb{E}[Z_{k-1}|\vth < Z_{k-1}, \UpdateKMO] \\
		& = \sum_{c=\vfloor - \tkmo}^{\Txcount} (1+c+\tkmo) 
		\frac{\OutageProb{k-1}^{c-1}\TxProbxTime{k-1}{1}}{\OutageProb{k-1}^{\vfloor -1 - \tkmo} - \OutageProb{k-1}^{\Txcount}}  \\
		&= 1 + \tkmo + \frac{1}{\TxProbxTime{k-1}{1}} \\
		&+ \frac{(\vfloor -1 - \tkmo) \OutageProb{k-1}^{\vfloor -1 - \tkmo} - \Txcount \OutageProb{k-1}^{\Txcount}}
		{\OutageProb{k-1}^{\vfloor -1 - \tkmo} - \OutageProb{k-1}^{\Txcount}}.
	\end{aligned}
\end{align}
By substituting \eqref{eq:Case2_P} and \eqref{eq:Case2_ZvleqZ} into \eqref{eq:Case2_rewritten}, $G(\UpdateKMO, \UpdateK)$ can be obtained where the result is given in \eqref{eq:gk_cndt_result_onecol}, case $v_1 \leq \vth < v_2$.

For $1 + \Txcount + \tkmo \leq \vth < \RoundLen + 2 + \tk$, we have
\begin{align}\label{eq:Case3_P}
	\POne = 0, \PTwo = 1, \PThree = 0.
\end{align}
By substituting \eqref{eq:Case3_P} into \eqref{eq:gk_cndt_def}, $G(\UpdateKMO, \UpdateK)$ can be obtained where the result is given in \eqref{eq:gk_cndt_result_onecol}, case $v_2 \leq \vth < v_{3,y=1}$.
%

For $\y\RoundLen+2+\tk \leq \vth < \y\RoundLen + 1 +\Txcount+\tk,\; \y=1,2,\cdots,$
$\vth$ is always larger than $Z_{k-1}$, i.e., $\POne = 0$. 
In addition, $\AVPCaseThree$ is satisfied when the update succeeds within $(\y-1)$ rounds or within $\FuncThree{k}{\y}$ slots at $\y$-th round, 
where $\FuncThree{k}{\y}$ is defined in \eqref{eq:FuncThree}.
Hence, we have
\begin{align}\label{eq:Case4_P}
	\begin{aligned}
		\POne &= 0,\\
		\PThree &=  1 + (1 - \Px)^{\y-1} \left(\Px \frac{\TxProbxTime{k}{\FuncThree{k}{\y}}}{\TxProbPf{k}} - 1 \right),\\
		\PTwo &= 1 - \PThree \\
		&= - (1 - \Px)^{\y-1} \left(\Px \frac{\TxProbxTime{k}{\FuncThree{k}{\y}}}{\TxProbPf{k}} - 1 \right).
	\end{aligned}
\end{align}
By substituting \eqref{eq:Case5_P} into \eqref{eq:gk_cndt_def},  $G(\UpdateKMO, \UpdateK)$ is given as
\begin{align}\label{eq:Case4_result}
	\begin{aligned}
		&G(\UpdateKMO, \UpdateK) \\
		&= \PTwo \mathbb{E}[\RoundLen + X_k - (\vfloor+1)|\AVPCaseTwo] \\
		&= \Px (1-\Px)^{\y-1} 
		\sum_{c=\FuncThree{k}{\y} + 1}^{\Txcount} \{(\y-1)\RoundLen + 1+c+\tk\} \\
		& \times \frac{\OutageProb{k}^{c-1} \TxProbxTime{k}{1}}{\TxProbPf{k}}
		+ \sum_{\Y=\y}^{\infty} \Px (1-\Px)^{\Y} \{\Y\RoundLen + \mathbb{E}[Z_k|\UpdateK] \} \\
		&- (1 - \Px)^{\y-1} \left(\Px \frac{\TxProbxTime{k}{\FuncThree{k}{\y}}}{\TxProbPf{k}} - 1 \right)
		\{\RoundLen - (\vfloor + 1)\},
	\end{aligned}
\end{align}
where the result is given in \eqref{eq:gk_cndt_result_onecol}, case $v_{3}(\y) \leq \vth < v_{4}(\y)$.

For $\y\RoundLen + 1 +\Txcount+\tk \leq \vth < (\y+1) \RoundLen + 2 + \tk,\; \y=1,2,\cdots$,
$\POne = 0$, similar to the above case.
In addition, $\AVPCaseThree$ is satisfied when the update succeeds within $\y$ rounds, so we have
\begin{align}\label{eq:Case5_P}
	\begin{aligned}
		\POne &= 0, \quad
		\PThree = \sum_{\Y=1}^{\y} \Px (1 - \Px)^{\Y-1} = 1 - (1 - \Px)^{\y},\\
		\PTwo &= 1 - \PThree = (1 - \Px)^{\y}.
	\end{aligned}
\end{align}
By substituting \eqref{eq:Case5_P} into \eqref{eq:gk_cndt_def},  $G(\UpdateKMO, \UpdateK)$ is obtained as
\begin{align}\label{eq:Case5_result}
	\begin{aligned}
		&G(\UpdateKMO, \UpdateK) \\
		&=\sum_{\Y=\y+1}^{\infty} \Px (1 - \Px)^{\Y-1}
		\{(\Y-1)\RoundLen + \mathbb{E}[Z_k|\UpdateK] \}\\
		&+ (1 - \Px)^{\y}\{\RoundLen - (\vfloor + 1)\},
	\end{aligned}
\end{align}
where the result is given in \eqref{eq:gk_cndt_result_onecol}, case $v_{4}(\y) \leq \vth < v_{3}(\y+1)$.

From \eqref{eq:gk_cndt_result_onecol}, the expected violation time $\gk$ can be obtained by \eqref{eq:E_gk_def}.
Finally, from \eqref{eq:AVP}, the \AVP\ can be expressed by dividing \blue{the obtained} $\gk$ into $\Uk$.

\end{appendix}

\bibliographystyle{IEEEtran}
\bibliography{Bibtex/IEEEabrv,Bibtex/ISCGroup, Bibtex/StringDefinitions}

\newpage

\end{document}